\documentclass[useAMS,usenatbib]{mnras}

\usepackage[active]{srcltx}
\usepackage{graphicx}
\usepackage{txfonts}
\usepackage{natbib}
\usepackage{multirow}
\usepackage{longtable}
\usepackage[applemac]{inputenc}
\usepackage{rotating}
\usepackage{tablefootnote}
\usepackage{url}
\usepackage[multiple]{footmisc}

\usepackage{color}
\usepackage{soul}
\definecolor{jaune}{rgb}{1.0, 1.0, 0.0}

\newcommand{\HD}{HD\,148937}

\newcommand{\kms}{km\,s$^{-1}\,$}

\def\gtrsim{\mathrel{\hbox{\rlap{\hbox{\lower4pt\hbox{$\sim$}}}\hbox{$>$}}}}
\def\ltsim{\mathrel{\hbox{\rlap{\hbox{\lower4pt\hbox{$\sim$}}}\hbox{$<$}}}}

\title[\HD]{A remarkable change of the spectrum of the magnetic Of?p star \HD\ reveals evidence of an eccentric, high-mass binary}

\author[\HD:\,an eccentric, high-mass binary]{G.A. Wade\thanks{E-mail: wade-g@rmc.ca}$^1$, J.V. Smoker$^2$, C.J. Evans$^3$, I.D. Howarth$^4$, R. Barba$^5$, N.L.J. Cox$^{6,7}$,   
\newauthor{N. Morrell$^8$, Y. Naz\'e$^{9}$, J. Cami$^{10,11}$, A. Farhang$^{10,12}$, N.R. Walborn$^{13}$\thanks{Deceased 22 February, 2018.}, J. Arias$^{5}$, R. Gamen$^{14}$}
\\
$^{1}$Department of Physics \& Space Science, Royal Military College of Canada, PO Box 17000 Station Forces, Kingston, ON, Canada K7K 0C6 \\
$^2$European Southern Observatory, Alonso de Cordova 3107, Casilla 19001, Vitacura, Santiago 19, Chile\\
$^{3}$UK Astronomy Technology Centre, Royal Observatory, Blackford Hill, Edinburgh, EH9 3HJ, UK\\
$^4$Department of Physics \& Astronomy, University College London, Gower Street, London WC1E 6BT, UK \\ 
$^5$Departamento de F\'isica y Astronom\'ia, Universidad de La Serena, Av. Cisternas 1200, La Serena, Chile\\
$^{6}$Anton Pannekoek Institute for Astronomy, University of Amsterdam, NL-1090 GE Amsterdam, The Netherlands\\
$^{7}$ACRI-ST, 260 Route du Pin Montard, Sophia Antipolis, France\\
$^8$Las Campanas Observatory, Carnegie Observatories, Casilla 601, La Serena, Chile\\
$^{9}$FNRS - University of Li\`ege, B5C, All\'ee du 6 Ao\^ut 19c, B4000-Li\`ege, Belgium\\
$^{10}$Department of Physics and Astronomy and Centre for Planetary
  Science and Exploration (CPSX), The University of Western Ontario,
  London, ON N6A 3K7, Canada\\
$^{11}$SETI Institute, 189 Bernardo Ave, Suite 100, Mountain View, CA 94043, USA\\ 
$^{12}$School of Astronomy, Institute for Research in Fundamental Sciences, 19395-5531 Tehran, Iran\\
$^{13}$Space Telescope Science Institute, 3700 San Martin Drive, Baltimore, MD 21218, USA\\
$^{14}$Instituto de Astrof\'isica de La Plata, CONICET--UNLP and Facultad de Ciencias Astron\'omicas y Geof\'isicas, UNLP, Paseo del Bosque, s/n, La Plata, Argentina}

\begin{document}

\date{Accepted . Received , in original form }

\pagerange{\pageref{firstpage}--\pageref{lastpage}} \pubyear{2002}

\maketitle

\label{firstpage}

\begin{abstract}
We report new spectroscopic observations of the magnetic Of?p star \HD\ obtained since 2015 that differ qualitatively from its extensive historical record of weak, periodic spectral variations. This remarkable behaviour represents clear evidence for an unprecedented change in the character of variability of the star. In this paper we describe the new spectral properties and compare them to the previous line profiles. Based on measurements of the radial velocities of the C~{\sc iii}/N~{\sc iii} emission lines near 4640~\AA\ and the C~{\sc iv} absorption lines near 5800~\AA, we infer that \HD\ is likely a high-mass, double-lined spectroscopic binary. Combining the spectroscopic orbit with an archival interferometric measurement of the apparent separation of the equal-brightness components, we tentatively conclude that \HD\ consists of two O-type stars with masses of approximately $34$ and $49~M_\odot$, orbiting in an eccentric ($e=0.75$), long-period ($P_{\rm orb}\sim 26$~y) orbit. We discuss the potential relationship of the binary system to the peculiar properties of \HD, and propose future observations to refine the orbital and stellar properties.
 \end{abstract}


\section{Introduction}

Magnetism in O-type stars is a recently-discovered phenomenon: the first magnetic O-type star was
reported only in 2002 \citep{2002MNRAS.333...55D}, and since that time
large-scale surveys such as the CFHT's MiMeS \citep{2016MNRAS.456....2W,2017MNRAS.465.2432G} and ESO's BOB 
\citep{2015IAUS..307..342M} Large Programs have
succeeded in identifying only a dozen in our Galaxy. As a consequence of the small sample size,
our understanding of their properties is limited. The known population \citep[e.g.][]{2015ASPC..494...30W}
ranges in spectral type from around O6 to O9.5. Although spectral types are variable and luminosity classes 
challenging to establish for many magnetic O stars, most appear to be main sequence objects. The detected magnetic fields are
generally oblique dipoles, with polar strengths ranging from $\sim 0.1$~kG to over 20~kG.
Their magnetic fields channel their powerful winds, resulting in dense circumstellar magnetospheres
confined to co-rotate with the star \citep{petit2013}. As a consequence, magnetic O stars exhibit strong variability across the electromagnetic
spectrum, with line and continuum emission modulated periodically according to the stellar rotation period.

Notwithstanding their relative rarity (\citealt{2017MNRAS.465.2432G} reported that just $7\pm 3$\% of O stars observed in the MiMeS survey host detectable magnetic fields), recent work has shown that magnetic O stars may provide new keys to resolving a number
of important, outstanding puzzles in stellar evolution, including the
large component masses of some merging black holes in binary systems \citep{2017MNRAS.466.1052P}, 
and the occurrence of pair-instability supernovae in
high-metallicity environments \citep{2017A&A...599L...5G}.

\HD\ was classified as a member of the peculiar Of?p spectral class by \citet{1972AJ.....77..312W,1973AJ.....78.1067W}.
The spectral peculiarities that are diagnostic of this class are now known to be intimately associated with magnetism, and
magnetic fields have been detected in all known Galactic members \citep{2017MNRAS.465.2432G}.
\citet{2008AJ....135.1946N,2010A&A...520A..59N} reported variability of optical spectral lines of \HD\ with 
a period of 7.03\,d, which they speculated may be related to the presence of a magnetic field, with the star viewed near the rotational pole.

At the time of its identification as a magnetic star in 2011 \citep{2011A&A...528A.151H,2012MNRAS.419.2459W}, 
\HD\ was only the fifth known magnetic O star. It remains both
the hottest and most massive star of this class \citep[$T_{\rm eff}=41$~kK, $M\sim 60~M_\odot$;][but see further discussion in Sect. 5 of this paper]{2008AJ....135.1946N,2012MNRAS.419.2459W}. ESPaDOnS \citep[e.g.][]{2012MNRAS.426.1003S} Stokes $I$ and $V$ observations carried out by
the MiMeS collaboration \citep{2012MNRAS.419.2459W} detected a weak ($\sim 300$~G) longitudinal magnetic field and
confirmed relatively subtle variability of both the longitudinal
field and many optical spectral lines according to the $7.03$\,d
period. \citet{2012MNRAS.419.2459W}
interpreted the results in the context of the Oblique Rotator Model \citep[ORM;][]{1950MNRAS.110..395S},
inferring that the star hosts an oblique, dipolar magnetic field that
confines the star's outflowing wind into a co-rotating dynamical magnetosphere with a magnetic wind confinement parameter (which measures the degree to which the stellar wind dynamics are influenced by the magnetic field) equal to $\eta_*=20$.
As a consequence, the periodic variability observed in the spectroscopic record is interpreted as rotational modulation according to its 7.03\,d
rotation period.

\begin{figure*}
\begin{tabular}{ccc}
\includegraphics[width=8.8cm]{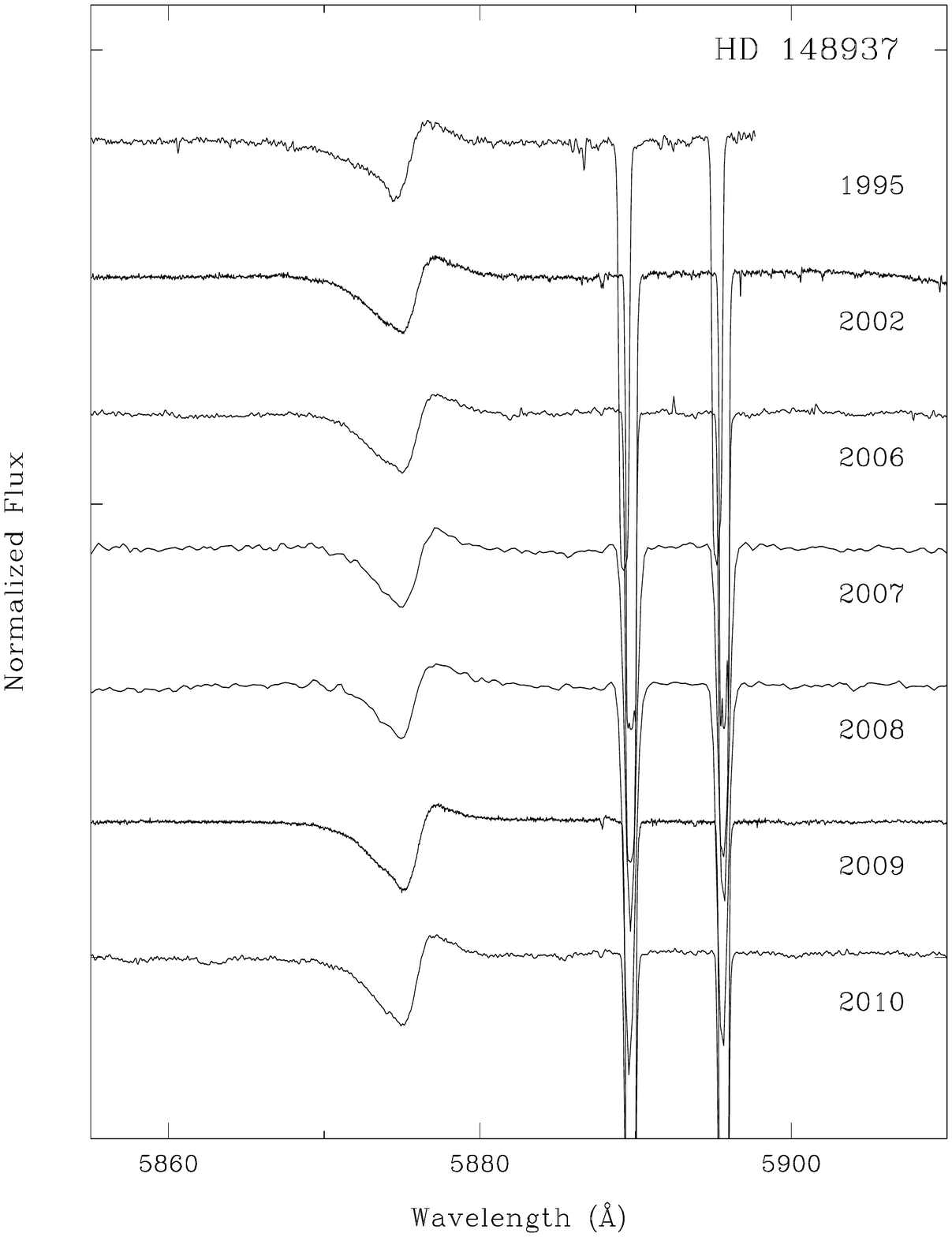}\includegraphics[width=8.8cm]{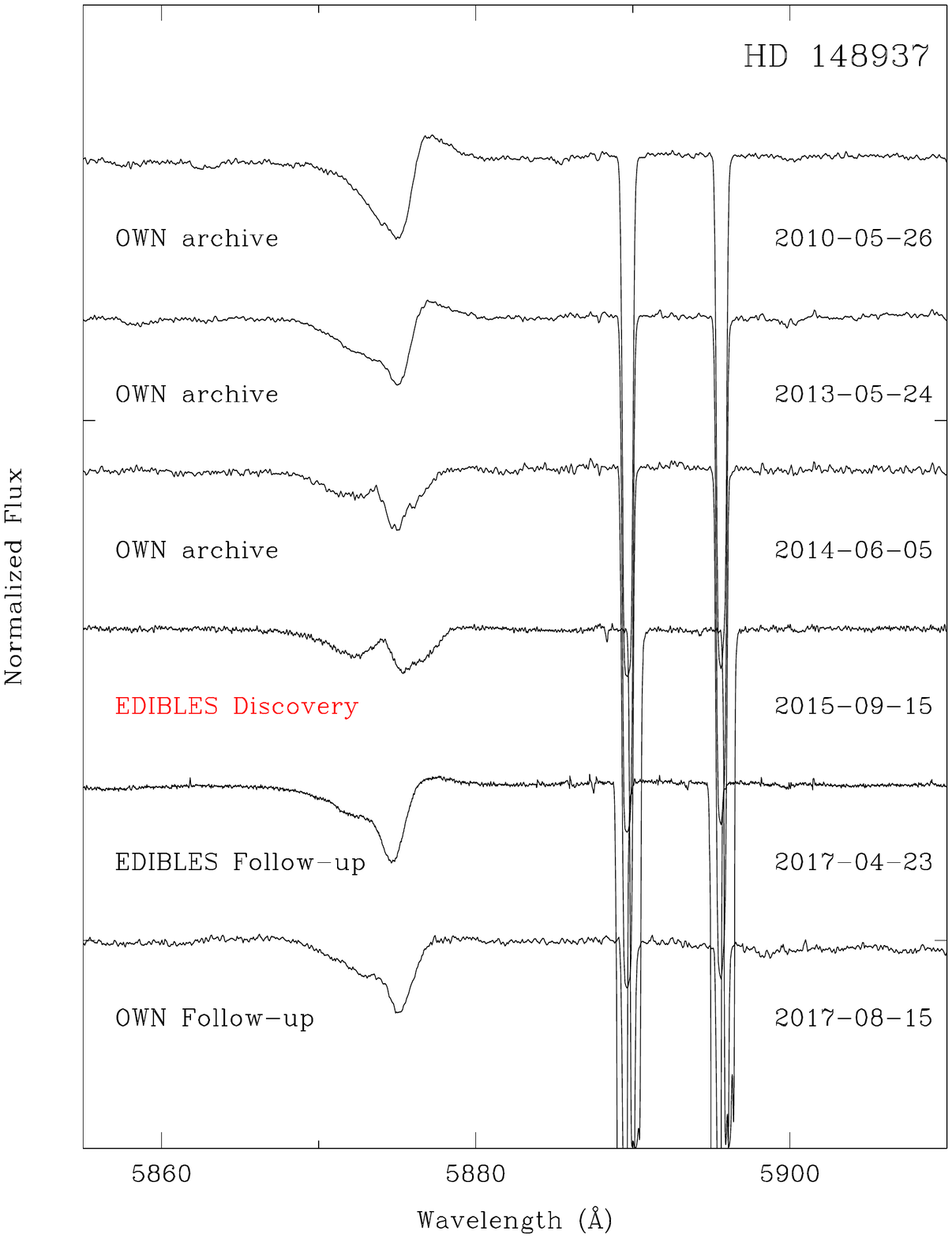}  \\
\end{tabular}
\caption{\label{5876}{\em Left -}\ Illustration of the telluric-corrected line profiles of He~{\sc i}~$\lambda 5876$ of \HD\ prior to 2011. All profiles except the 1995 (UCLES) observation are qualitatively consistent with the range of profile shapes attributed to rotational modulation by \citet{2012MNRAS.419.2459W}. The UCLES profile shows a weaker extended blue wing that distinguishes it, and that may indicate that another episode, similar to that inferred to be occurring now, was underway at that time. {\em Right -}\ Illustration of the change in the character of the He~{\sc i}~$\lambda 5876$ profile in 2013, and its subsequent evolution.}
\end{figure*}

Many spectral lines of \HD\ -- including lines in strong emission, as well as those in absorption -- exhibit detectable variability according to this period (\citealt{2010A&A...520A..59N}; see Fig. 5 of \citealt{2012MNRAS.419.2459W}). The measured equivalent widths (EWs) of the spectral lines vary approximately sinusoidally, and emission and absorption lines vary approximately in phase. This variability is also explained by the ORM, in which rotation of the star and its oblique magnetosphere results in periodic modulation of the spectrum. For \HD, \citet{2012MNRAS.419.2459W} inferred a rotation axis inclination of $i\leq 30\degr$, a magnetic obliquity $\beta\sim 40\degr$, and a magnetic dipole strength of $B_{\rm d}$ of 1~kG. 

Early studies left a number of peculiar characteristics of this system
unexplained. In particular, \citet{2012MNRAS.419.2459W}  commented that the star exhibits
large cycle-to-cycle changes in the line profiles that probe the
magnetosphere. They attributed these variations to ``intrinsic changes in the 
amount or distribution of emitting material with time, i.e. evolution of the magnetosphere of 
HD 148937''. However, they were unable to determine if the proposed evolution was secular or stochastic. 
They also noted that while the Zeeman signatures of HD 148937
observed in mean (Least Squares Deconvolved, or LSD) line profiles were very weak, a large Stokes $V$
signal was observed in the absorption component of the He~{\sc i} $\lambda 5876$ line.
Since this line is not a particularly sensitive magnetic diagnostic,
the appearance of a strong Zeeman signature in this line (while being absent from other lines of comparable magnetic sensitivity) was not straightforwardly
understood by \citet{2012MNRAS.419.2459W}. 

Finally, \HD\ is distinguished from other magnetic O stars by its remarkable bipolar
nebula, NGC 6164/5 \citep[see][and references therein]{2017A&A...599A..61M}. 
Recently, \citet{2017A&A...599A..61M} analyzed Spitzer and Herschel
observations of the NGC 6164/5 nebula to constrain its global
morphology, kinematics and abundances, and
the evolutionary status of the central star. They concluded that the
structure of the nebula is particularly complex and is composed of a
close bipolar ejecta nebula, an ellipsoidal wind-blown
shell, and a spherically symmetric Str\"omgren sphere. The combined
analyses of the known kinematics and of the new abundances of the
nebula suggest either a helical morphology for the nebula, possibly
linked to the magnetic geometry, or the occurrence of a binary merger.
The latter hypothesis would be of great interest, since binary mergers
have been proposed as a potential pathway to the
generation of magnetic fields in massive stars \citep[e.g.][]{2012ARA&A..50..107L} .


\HD\ was resolved as an interferometric binary in the course of the Southern Massive Stars at High angular resolution survey \citep[SMASH+;][]{2014ApJS..215...15S}. {Those authors reported that \HD\ is an equal-brightness stellar pair ($\Delta m=0.00\pm 0.02$ in $H$-band) with separation $\rho = 21.05\pm 0.67$~mas. {\citet{2014ApJS..215...15S} further detected a faint ($\Delta H = 5.39\pm 0.15$) companion at $3.30\pm 0.06$~arcsec.} Interestingly, none of the previous spectroscopic and imaging studies of the system \citep{2008AJ....135.1946N,2010A&A...520A..59N,2012MNRAS.419.2459W,2017A&A...599A..61M} have reported any direct evidence of binarity. 

The measured angular separation can be leveraged to determine the instantaneous projected physical separation of the equal-brightness pair. The revised Hipparcos parallax implies a distance $d=0.43\pm 0.11$~kpc, yielding a separation of 8.9 AU\footnote{We note that the paper by \citet{{2014ApJS..215...15S}} appears to contain a typo, since those authors report an instantaneous projected separation of 40~AU based on the Hipparcos distance. We note furthermore that the article reports two slightly different values of $\rho$. H. Sana (priv. comm.) confirms that $\rho=21.05\pm 0.67$~mas as reported in table 1 of that paper is in fact the correct value.}}. On the other hand, the Gaia DR2 parallax \citep{2018arXiv180409366L} yields a significantly greater distance of $1.14\pm 0.06$~kpc, which implies an instantaneous separation nearly three times larger. However, both results may be compromised by the binarity (see Sect.~6.1).


In this paper we report a significant, unprecedented change in the optical line profiles of \HD\ that stands in stark contrast to the historical record of its periodic variability. Potential origins of the unexpected variability could be associated with the probable binarity detected by interferometry. A companion of comparable mass and luminosity to the magnetic star might, for example, also be detectable spectroscopically, and its orbital motion might translate into spectral variability. Alternatively, the two stars might interact episodically or periodically, or one of the components might undergo instability and eruption, e.g. a Luminous Blue Variable (LBV)-like eruption. Another more exotic possibility is a significant reconfiguration of the stellar magnetosphere. 

In order to better understand the origin of the change of the spectrum of \HD, we describe and analyze new spectroscopic observations which, in combination with archival data, provide new insight into the character of the \HD\ system.

\section{Observations}\label{obs}


\subsection{Optical spectroscopy}

High-resolution, high signal-to-noise ratio (S/N) spectra of \HD\ were obtained in August 2015, September 2016, April 2017, May 2017, and June 2017
in the context of the ESO Diffuse Interstellar Band Large Exploration Survey (EDIBLES) Large Program \citep{2017A&A...606A..76C}. 

\HD\ was observed with the UVES spectrograph \citep{2000SPIE.4008..534D}  mounted on the Very Large Telescope (VLT)
at the Paranal Observatory. UVES covers
the optical spectrum from 305~nm to 1042~nm at high spectral
resolving power ($R\sim$~71,000 in the near-UV/blue to $R\sim$~107,000 in
the red part of the spectrum). The scientific goals of the EDIBLES
program require a very high S/N for each of
the target spectra; typically the S/N is 700--1000 depending on the
wavelength range. To achieve this data quality, additional calibration
files and data reduction steps are necessary; see Cox et al. (2017)
for details.  \HD\ was observed in total on 5 different occasions by EDIBLES; however, on
each visit only part of the spectrum was observed. 

Additional spectroscopic observations of \HD\ were obtained during the intensive high-resolution spectroscopic campaign of Southern Galactic O- and WN-type stars \citep[the ``OWN Survey";][]{2010RMxAC..38...30B,2017IAUS..329...89B}. The data were obtained using the REOSC Cassegrain spectrograph\footnote{On long-term loan from the University of Li\`ege.} in cross-dispersed mode attached to the ``Jorge Sahade'' 2.15 m telescope at the Complejo Astron\'omico El Leoncito (CASLEO, Argentina; $R=15,000$, $3600\leq \lambda\leq 6100$~\AA) and the \'echelle spectrograph at the 2.5 m du Pont telescope of Las Campanas Observatory (LCO), in Chile ($R=40,000$, $3450\leq \lambda\leq 9850$~\AA). Observations were collected between May 2006 and August 2017. 
For wavelength calibration of CASLEO and LCO spectroscopic observations, we obtained calibration lamp (Th-Ar) exposures immediately before or after each target integration, at the same telescope position. The spectra were processed and calibrated using standard IRAF\footnote{IRAF is distributed by the National Optical Astronomy Observatories,
    which are operated by the Association of Universities for Research
    in Astronomy, Inc., under cooperative agreement with the National
    Science Foundation.} routines. 
    
    One spectrum was obtained with the Magellan Inamori Kyocera double Echelle spectrograph, MIKE \citep{2003SPIE.4841.1694B} on the 6.5m Magellan II (Clay) telescope. HD 148937 was observed through a 0.7\arcsec\ slit yielding $R \sim 40,000$ in the blue side (3350-5000~\AA) and $R \sim 33,000$ in the red side (4900-9500 \AA). Both detectors were read out in 2x2 binning mode. The S/N is around 200 in the region containing He~{\sc i} $\lambda 5876$. The data were reduced with IRAF echelle routines.



Ten FEROS spectra, 5 of which were first discussed and measured by \citet{2008AJ....135.1946N}, were also included. We also include a single spectrum obtained using the UCLES spectrograph at the Anglo-Australian Telescope in 1995, that was also discussed and analyzed by those authors. 


Finally, we also included a single UVES-POP observation \citep{2003Msngr.114...10B} obtained
in 2002, and a subset of the ESPaDOnS spectra (obtained in 2009 and 2010) reported by \citet{2012MNRAS.419.2459W}.

We note that while some other spectra of \HD\ may exist in the archives, these data either do not add significantly (typically having been acquired at dates similar to the data in hand) or are lacking coverage of the key spectral features (i.e. the $\lambda 5800$ and $\lambda 4640$ features) used in our study.

All of the spectra used in this study are summarized in Table~\ref{observations}.

\subsection{X-ray flux}

New X-ray observations were acquired to investigate if the X-ray emission of \HD\ has been impacted by the phenomena responsible for the recent spectroscopic changes (as is further discussed in Sect.~6.3).

We obtained a new observation of \HD\ with Chandra (\citet{1996SPIE.2805....2W}; ObsID = 20148, 9ks duration) in June 2018. The source spectrum was extracted using ``specextract" under CIAO v4.9 in a circle centered on the SIMBAD coordinates of \HD\ and with 5 pixels radius. The background spectrum was extracted in the surrounding annulus with external radius of 15 pixels. Response matrices necessary for flux and energy calibration were calculated after extraction and the spectrum was grouped to reach a minimum of 15 counts per bin. To avoid pile-up, the data were obtained only for a subarray of ACIS (chip \#7) in TE (timed exposure) mode and using a frame time of 0.4\,s.


As appropriate for massive stars, the spectral fitting was done within Xspec using an absorbed optically-thin thermal emission model of the type $wabs\times phabs\times(apec_1 + apec_2)$, considering plasma abundances fixed to those of \citet{1989GeCoA..53..197A}. The first component ($wabs$) represents the interstellar absorption, which was fixed to the appropriate value (i.e. absorbing column $N_{\rm H}^{int}=4\times 10^{21}$\, cm$^{-2}$, see \citealt{2012ApJ...746..142N} and references therein). The second one ($phabs$, which has a single fitted parameter $N_{\rm H}^{add}$) accounts for potential circumstellar absorption. The last components ($apec_{1,2}$) represent the thermal emission by the hot plasma. Two components are needed to fit well the observed spectra (see Naze et al. 2012b) ; each one has two free parameters : the temperature ($kT$) and the normalization factor ($norm$) which is linked to the emission measure of the plasma. As in  \citet{2012ApJ...746..142N}, there are two solutions of similar quality ($\chi^2\sim 1$) : the "cool" solution has $N_{\rm H}^{add}=(0.34\pm0.08)\times 10^{22}\,$cm$^{-2}$, $kT_1=0.29\pm0.03$\,keV, $norm_1=(1.1\pm0.4)\times 10^{-2}$cm$^{-5}$ , $kT_2=2.02\pm0.09$\,keV, and $norm_2=(3.20\pm0.15)\times 10^{-3}$cm$^{-5}$; the "hot" one has $N_{\rm H}^add=(0.07\pm0.08)\times 10^{22}$cm$^{-2}$, $kT_1=0.74\pm0.03$\,keV, $norm_1=(2.0\pm0.4)\times 10^{-3}$cm$^{-5}$ , $kT_2=2.6\pm0.2$\,keV, and $norm_2=(2.25\pm0.17)\times 10^{-3}$cm$^{-5}$. The observed flux in the 0.5--10.keV energy range is $(3.3\pm0.1)\times 10^{-12}$erg\,cm$^{-2}$\,s$^{-1}$.

\begin{table}
\begin{centering}
\begin{tabular}{rccrr}
\hline
HJD	&	UT date	&	Instrument	&	RV $\lambda 4640$			&	RV $\lambda 5800$			\\
-2400000	&		&		&	(km\,s$^{-1}\,$)			&	(km\,s$^{-1}\,$)			\\
\hline													
49822.5	&	       1995-04-15      	&	       UCLES$^*$       	&	$	-24.3	\pm	1	$	&	$	-23.9	\pm	3	$	\\
52332.841	&	       2002-02-27      	&	       UVES    	&	$	-15.5	\pm	1	$	&	$				$	\\
52783.757	&	       2003-05-24      	&	       FEROS	&	$	-17.6	\pm	1	$	&	$	-22.5	\pm	3	$	\\
52784.749	&	       2003-05-25      	&	       FEROS	&	$	-17.8	\pm	1	$	&	$	-21.4	\pm	3	$	\\
53131.618	&	       2004-05-06      	&	       FEROS   	&	$	-18.3	\pm	1	$	&	$	-25.0	\pm	3	$	\\
53509.728	&	       2005-05-19      	&	       FEROS   	&	$	-17.6	\pm	1	$	&	$	-23.8	\pm	3	$	\\
53547.535	&	       2005-06-26      	&	       FEROS   	&	$	-17.4	\pm	1	$	&	$	-28.2	\pm	3	$	\\
53866.8	&	       2006-05-11      	&	       CASLEO  	&	$	-12.4	\pm	3	$	&	$	-24.2	\pm	4	$	\\
53871.828	&	       2006-05-16      	&	       CASLEO  	&	$	-14.0	\pm	3	$	&	$	-32.3	\pm	5	$	\\
53874.758	&	       2006-05-19      	&	       du Pont echelle 	&	$	-19.6	\pm	1	$	&	$	-27.9	\pm	3	$	\\
53898.789	&	  2006-06-12      	&	       FEROS   	&	$	-17.4	\pm	1	$	&	$	-20.9	\pm	3	$	\\
53951.543	&	 2006-08-04      	&	       FEROS   	&	$	-17	\pm	1	$	&	$	-23.3	\pm	3	$	\\
53920.672	&	       2006-07-04      	&	       du Pont echelle 	&	$	-16.9	\pm	1	$	&	$	-30.4	\pm	3	$	\\
53938.632	&	       2006-07-22      	&	       du Pont echelle 	&	$	-18.4	\pm	1	$	&	$	-31.9	\pm	3	$	\\
53955.53	&	       2006-08-08      	&	       MIKE    	&	$	-18.6	\pm	1	$	&	$	-32.3	\pm	3	$	\\
54210.82	&	       2007-04-20      	&	       FEROS   	&	$	-18.3	\pm	3	$	&	$	-33.8	\pm	4	$	\\
54220.798	&	       2007-04-30      	&	       CASLEO  	&	$	-10.2	\pm	3	$	&	$	-28.1	\pm	4	$	\\
54252.711	&	       2007-06-01      	&	       CASLEO  	&	$	-4.6	\pm	3	$	&	$	-29.4	\pm	4	$	\\
54286.644	&	       2007-07-05      	&	       CASLEO  	&	$	-8.0	\pm	3	$	&	$	-33.7	\pm	4	$	\\
54511.873	&	       2008-02-15      	&	       CASLEO  	&	$	-11.1	\pm	3	$	&	$	-35.7	\pm	4	$	\\
54582.734	&	       2008-04-26      	&	       CASLEO  	&	$	-16.3	\pm	3	$	&	$	-30.1	\pm	4	$	\\
54608.737	&	       2008-05-22      	&	       CASLEO  	&	$	-13.3	\pm	3	$	&	$	-38.4	\pm	4	$	\\
54642.735	&	       2008-06-25      	&	       CASLEO  	&	$	-17.4	\pm	3	$	&	$	-28.1	\pm	4	$	\\
54659.564	&	       2008-07-12      	&	       CASLEO  	&	$	-17.6	\pm	3	$	&	$	-34.3	\pm	4	$	\\
54904.882	&	      2009-03-14       	&	      FEROS    	&	$	-17.8	\pm	2	$	&	$	-25.8	\pm	3	$	\\
54914.909	&	      2009-03-24       	&	      FEROS    	&	$	-18.5	\pm	2	$	&	$	-23.5	\pm	3	$	\\
54955.047	&	      2009-05-03      	&	 ESPaDOnS      	&	$	-17.2	\pm	1	$	&	$	-24.3	\pm	3	$	\\
55078.728	&	      2009-09-04      	&	 ESPaDOnS      	&	$	-18.9	\pm	1	$	&	$	-22.7	\pm	3	$	\\
55342.751	&	       2010-05-26      	&	       du Pont echelle 	&	$	-17.5	\pm	1	$	&	$	-32.0	\pm	3	$	\\
55400.771	&	       2010-07-23      	&	       ESPaDOnS        	&	$	-17.6	\pm	1	$	&	$	-25.3	\pm	2	$	\\
55403.778	&	       2010-07-26      	&	       ESPaDOnS        	&	$	-16.1	\pm	1	$	&	$	-23.7	\pm	2	$	\\
56436.738	&	       2013-05-24      	&	       du Pont echelle 	&	$	-26.1	\pm	1	$	&	$	-22.8	\pm	3	$	\\
56813.69	&	       2014-06-05      	&	       du Pont echelle 	&	$	-50.9	\pm	2	$	&	$	-6.3	\pm	3	$	\\
57251.589	&	       2015-08-17      	&	       UVES    	&	$	-55.1	\pm	1	$	&	$	-0.8	\pm	2	$	\\
57644.548	&	       2016-09-13      	&	       UVES    	&	$	-39.1	\pm	1	$	&                              \\			
57867.755	&	       2017-04-24      	&	       UVES    	&	$	-34.0	\pm	1	$	&	$	-18.5	\pm	2	$	\\
57890.906	&	      2017-05-17      	&	       UVES    	&	$	-35.9	\pm	1	$	&	                               \\			
57924.5	&	       2017-06-20      	&	       UVES    	&	$	-34.0	\pm	1	$	&		                             	\\
57933.636	&	       2017-06-29      	&	       du Pont echelle 	&	$	-31.6	\pm	1	$	&	$	-15.1	\pm	3	$	\\
57934.654	&	       2017-06-30      	&	       du Pont echelle 	&	$	-31.3	\pm	1	$	&	$	-17.7	\pm	3	$	\\
57935.654	&	       2017-07-01      	&	       du Pont echelle 	&	$	-31.1	\pm	1	$	&	$	-17.1	\pm	3	$	\\
57936.666	&	       2017-07-02      	&	       du Pont echelle 	&	$	-31.8	\pm	1	$	&	$	-16.2	\pm	3	$	\\
57980.614	&	       2017-08-15      	&	       du Pont echelle 	&	$	-31.7	\pm	1	$	&	$	-13.1	\pm	3	$	\\
57981.621	&	       2017-08-16      	&	       du Pont echelle 	&	$	-32.6	\pm	1	$	&	$	-15.1	\pm	3	$	\\
57983.574	&	       2017-08-18      	&	       du Pont echelle 	&	$	-31.8	\pm	1	$	&	$	-15.2	\pm	3	$	\\
\hline\hline
\end{tabular}
\end{centering}
\caption{Visible spectroscopy of \HD\ employed in this investigation and measurements of the heliocentric radial velocities of the N~{\sc iii}/C~{\sc iii} emission line complex at $\sim 4640$~\AA\ and the C~{\sc iv} absorption lines at $5801/11$~\AA. The single UCLES observation (marked with a $^*$), first reduced and analyzed by \citet{2008AJ....135.1946N}, has been re-reduced and re-measured for this study. }\label{observations}
\end{table}

%

\begin{figure*}
\begin{tabular}{ccc}
\vspace{-0.3cm}\includegraphics[width=6.25cm]{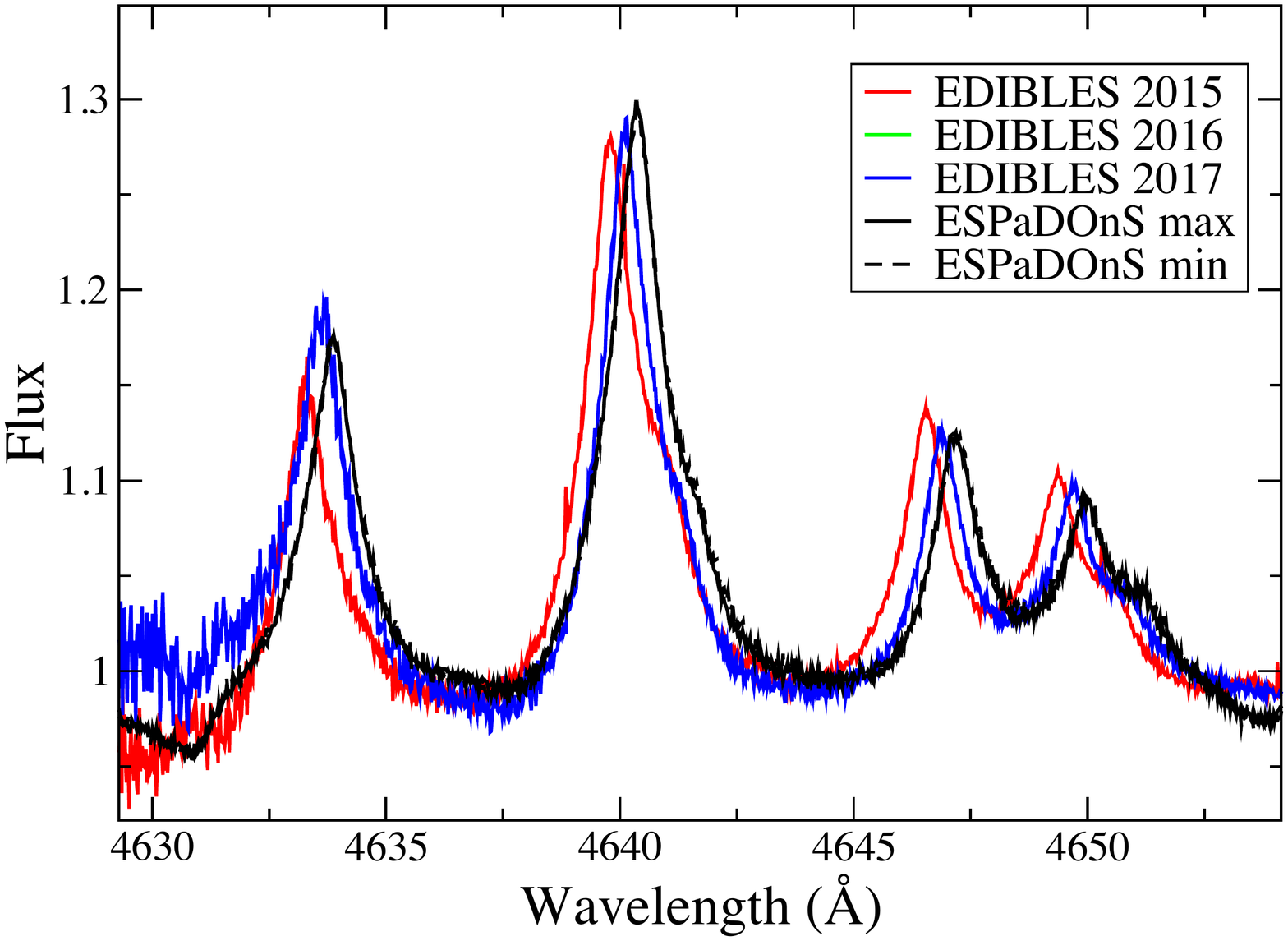}\hspace{-0.6cm}\includegraphics[width=6.25cm]{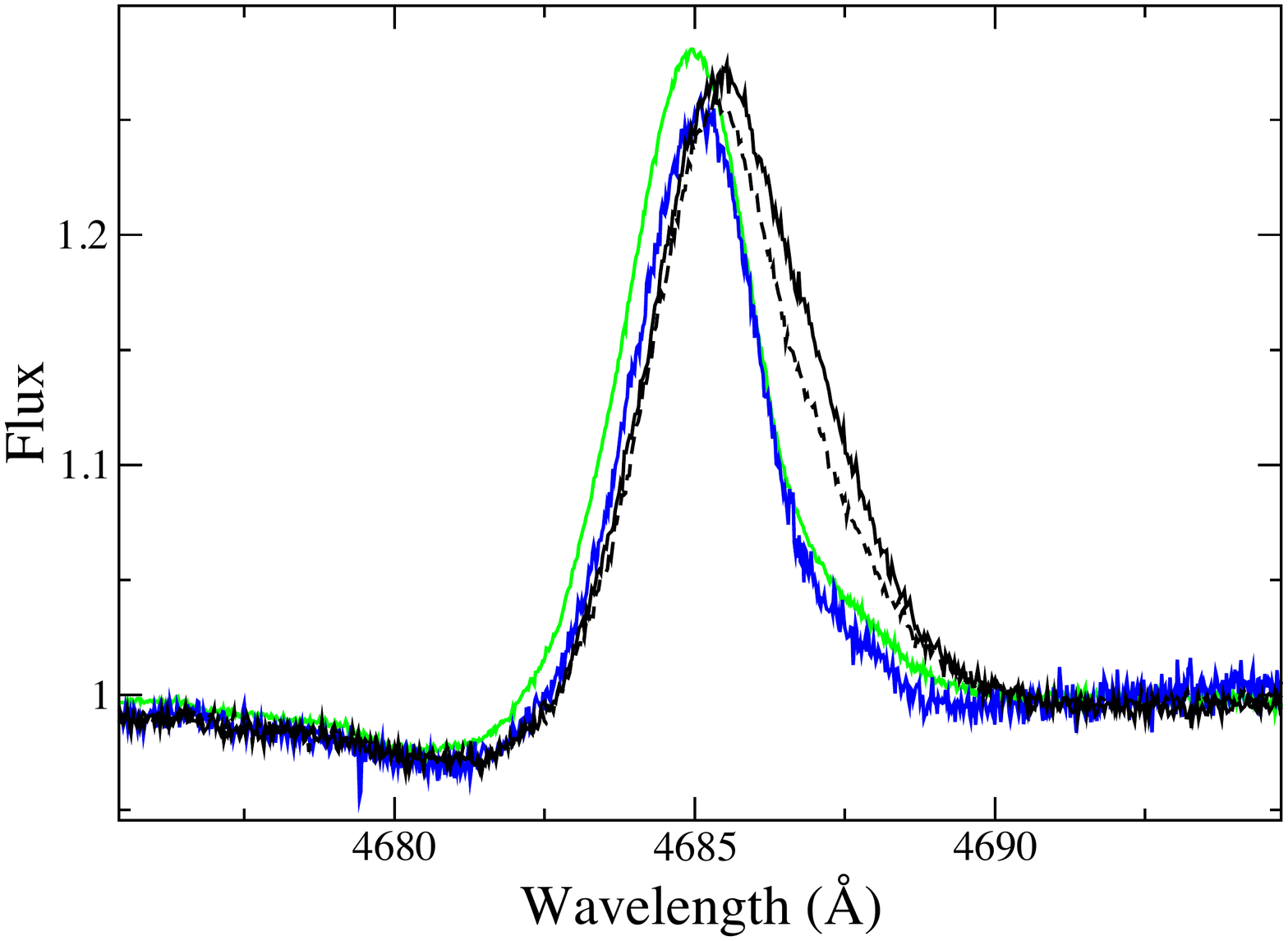}\hspace{-0.6cm}\includegraphics[width=6.25cm]{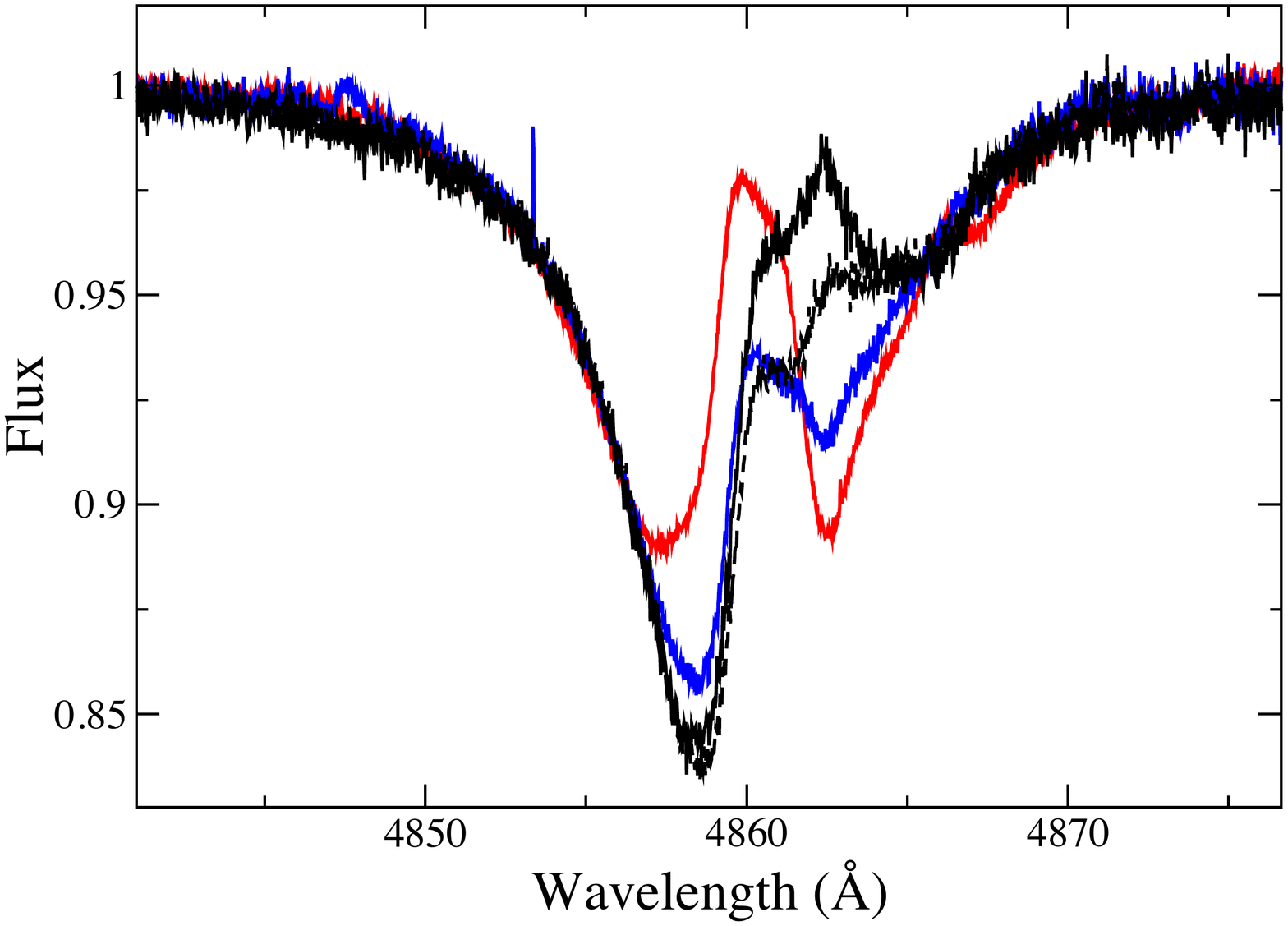}\\
\vspace{-0.3cm}\includegraphics[width=6.25cm]{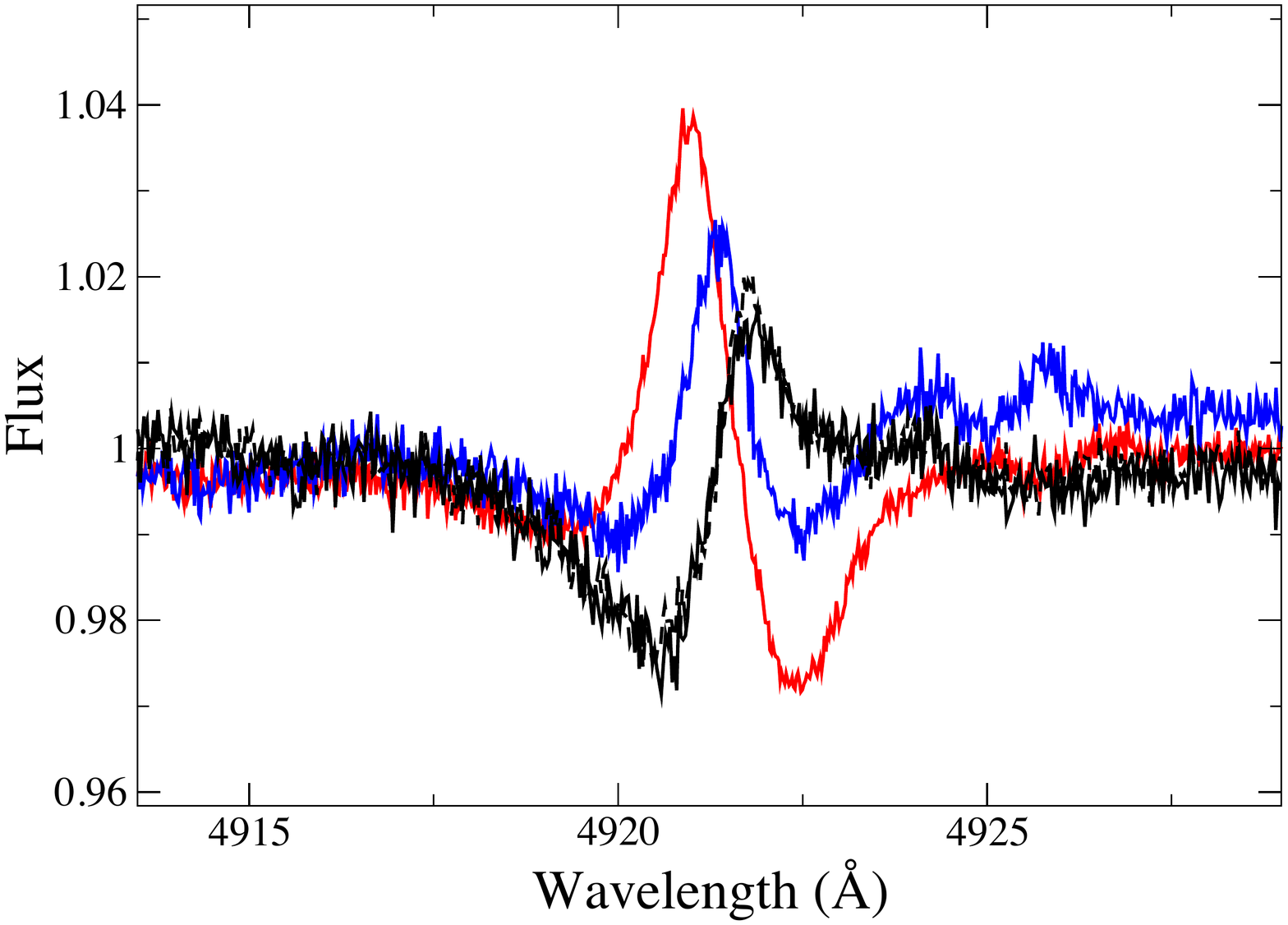}\hspace{-0.6cm}\includegraphics[width=6.25cm]{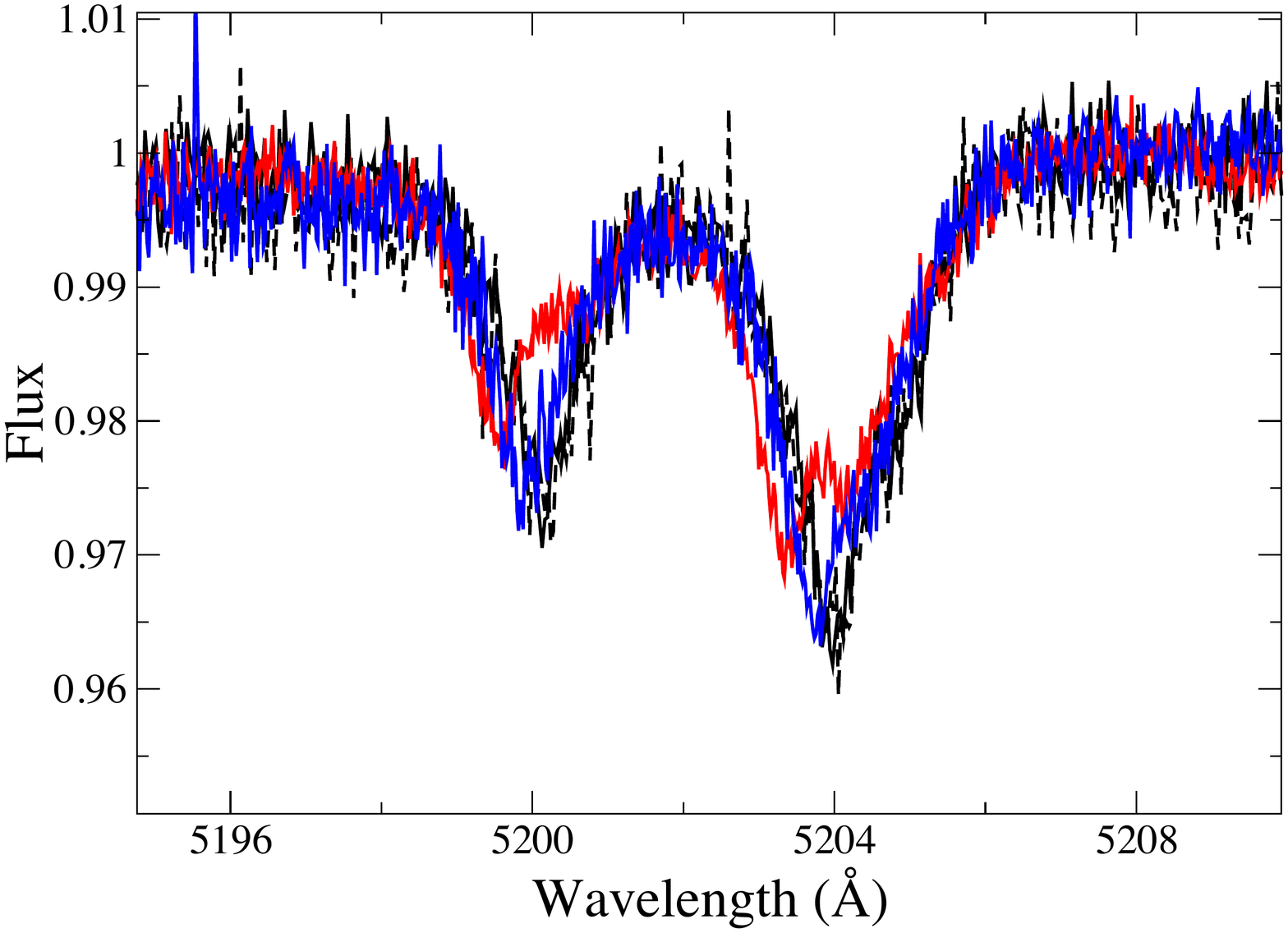}\hspace{-0.6cm}\includegraphics[width=6.25cm]{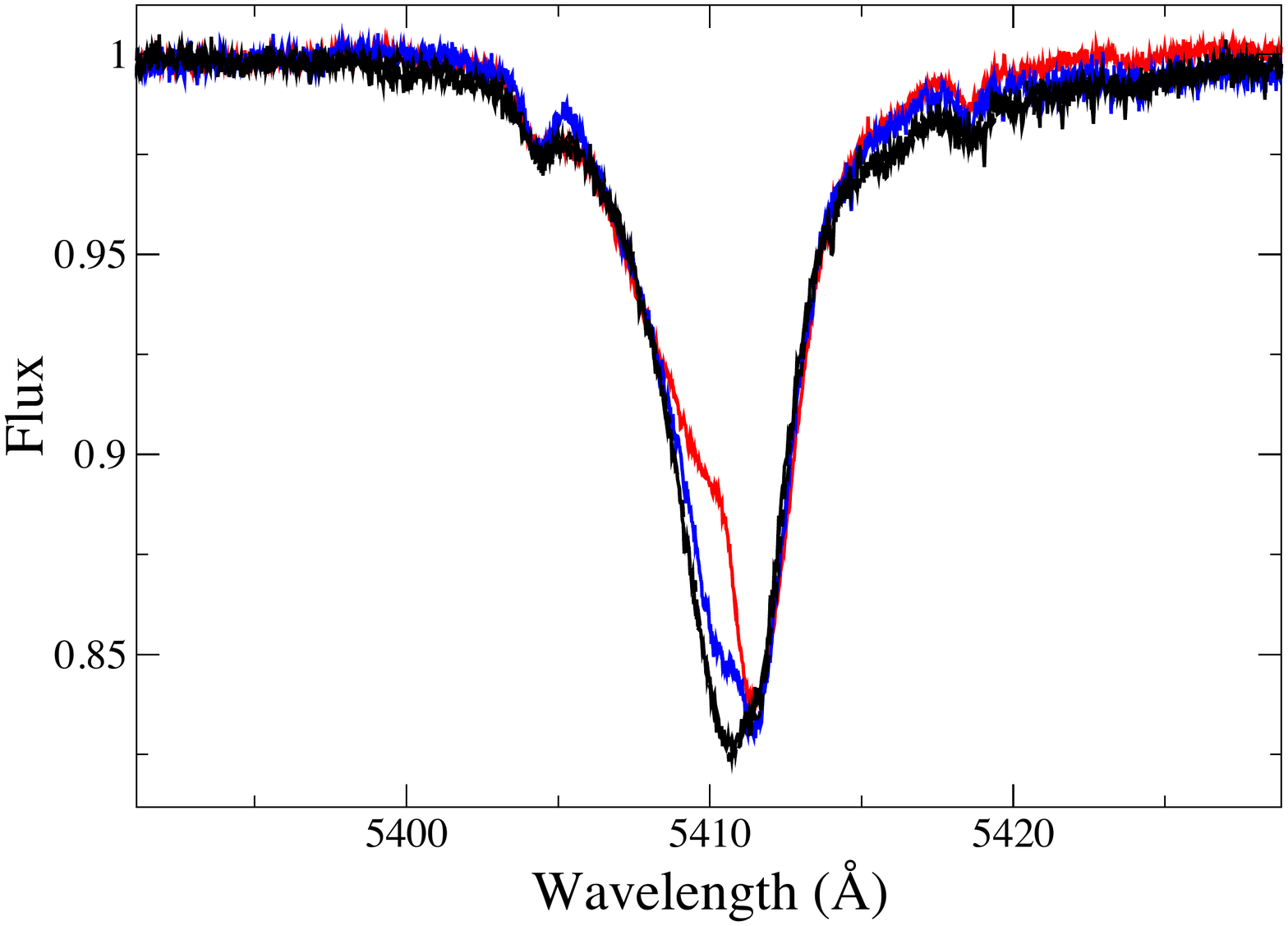}\\
\includegraphics[width=6.25cm]{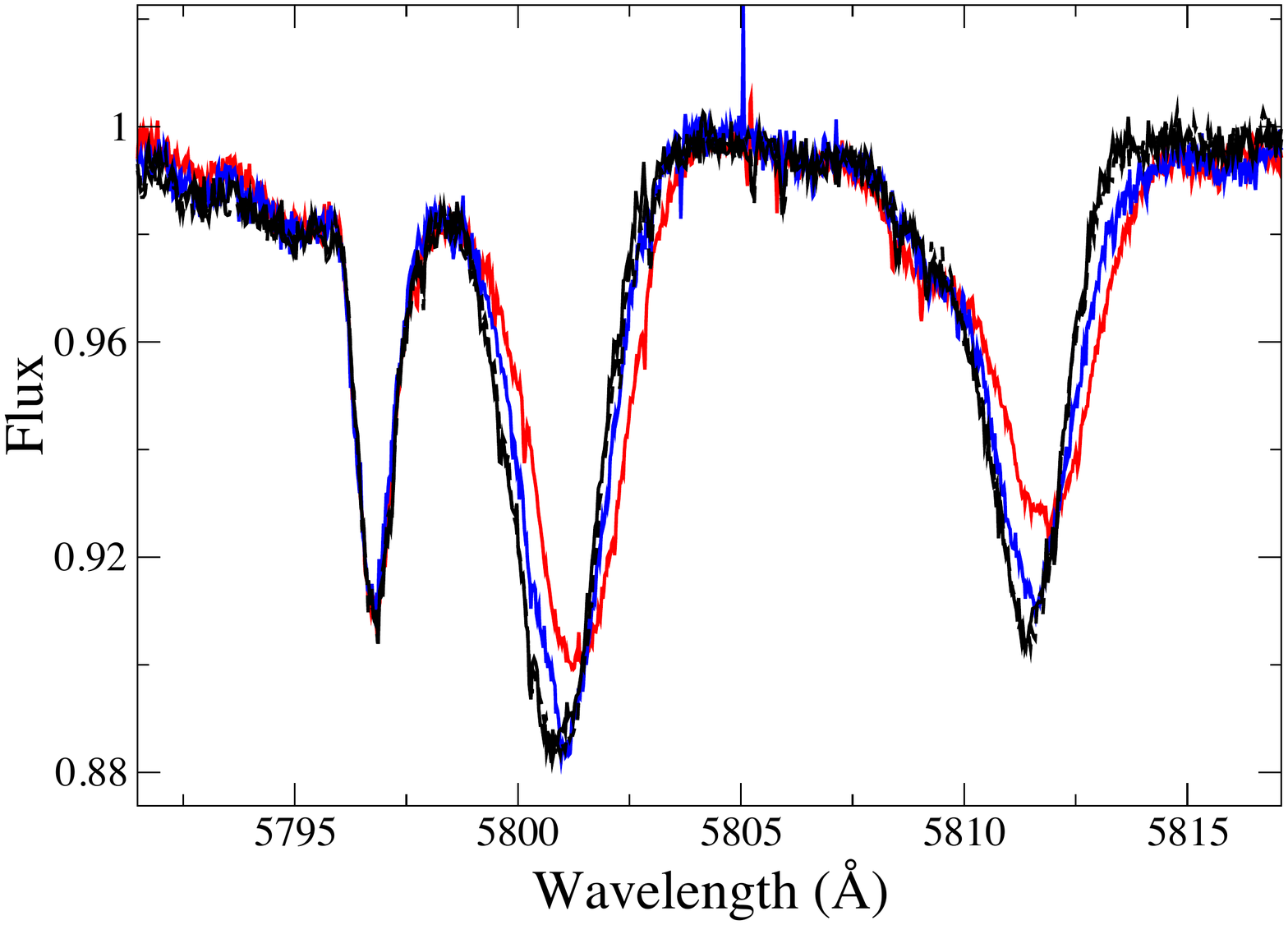}\hspace{-0.6cm}\includegraphics[width=6.25cm]{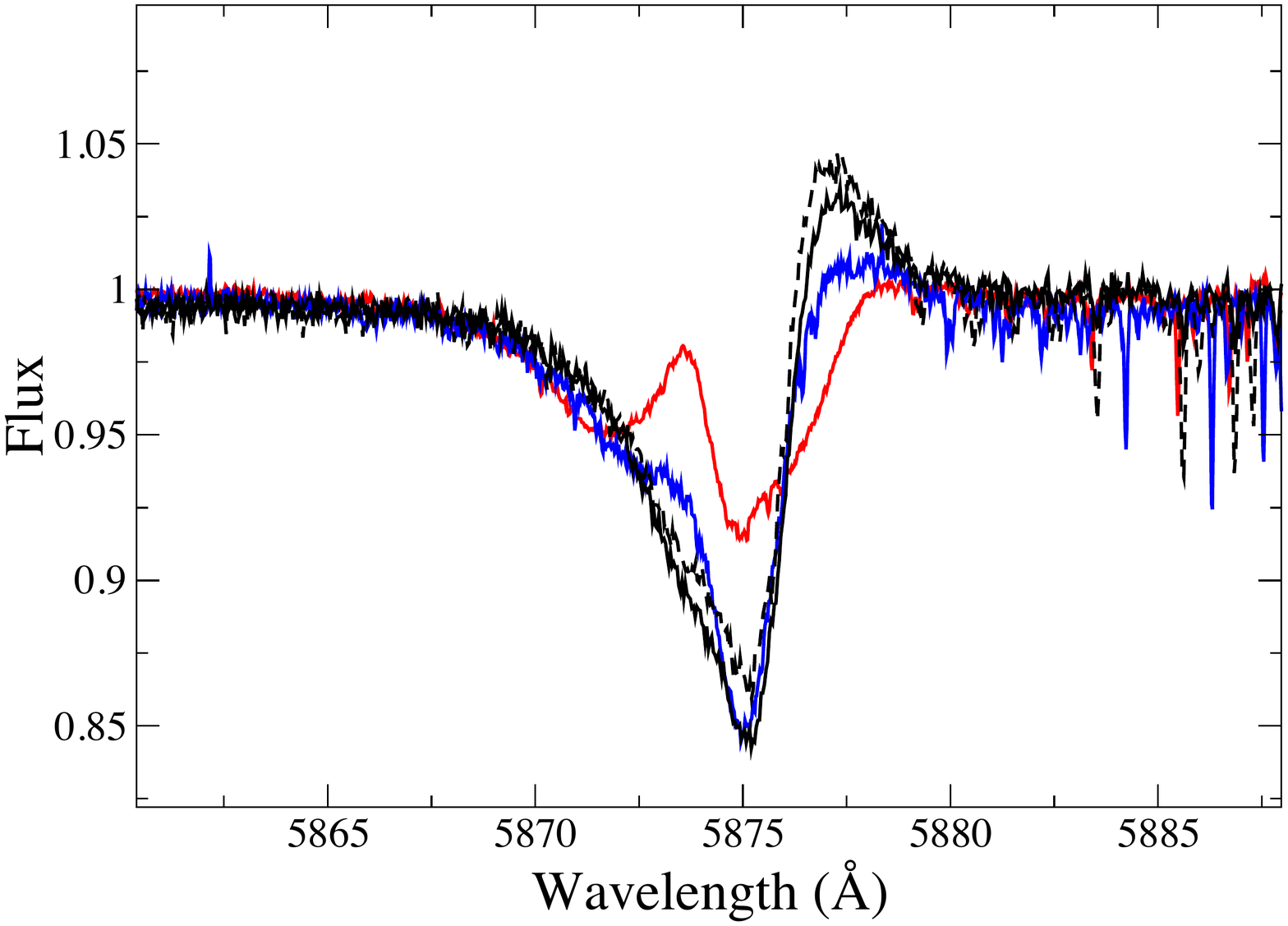}\hspace{-0.6cm}\includegraphics[width=6.25cm]{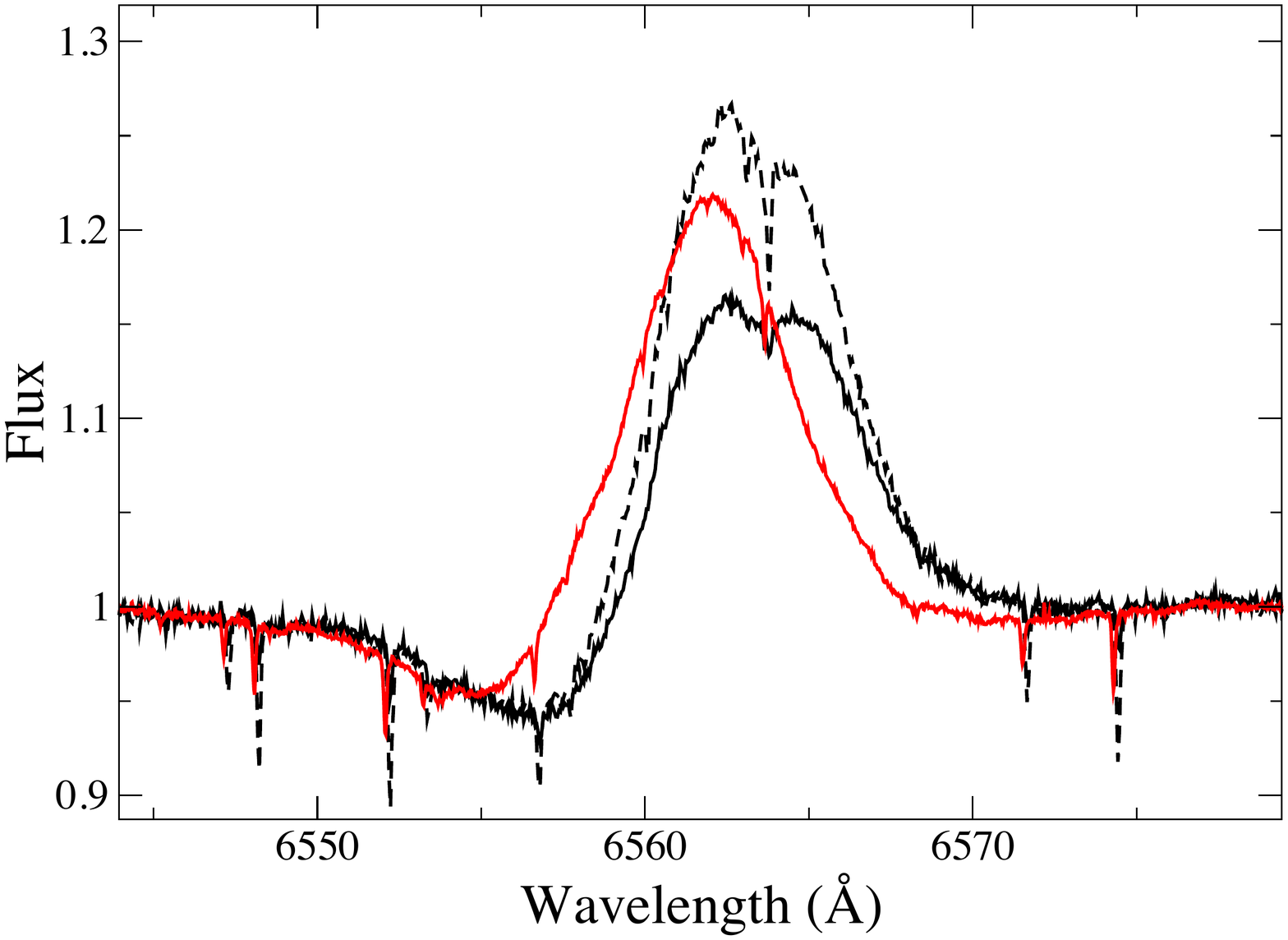}
\end{tabular}
\caption{\label{profiles}The black (solid and dashed) profiles show the range of variability of each spectral line in the ESPaDOnS dataset of Wade et al. (2012). The red, green and blue profiles show the EDIBLES observations obtained in 2015 (JD 2457251), 2016 (JD 2457644) and 2017 (JD 2457867), respectively. Spectra are in the heliocentric reference frame and have not been corrected for telluric absorption. }
\end{figure*}

\section{Long-term line profile variability of \HD}

The change in character of the spectrum of \HD\ was first identified by the EDIBLES team from profiles of the He~{\sc i}~$\lambda 5876$ line obtained in 2015 (on JD 2457251, which we will refer to as `the 2015 spectrum'). As illustrated in the left panel of Fig.~\ref{5876}, the profile of the $\lambda 5876$ line maintained the same essential quasi-P~Cyg morphology in observations obtained from 1995-2010. The (small) range of variability of the profiles acquired during this period is in good general agreement with the 7.03\,d rotational variability measured in past studies.

As shown in the right panel of Fig.~\ref{5876}, the 2015 UVES spectrum revealed a stark change in the $\lambda 5876$ profile; the initiation of this change can be traced back to at least 2013, and the profiles appear to be returning to normal in 2017 (see Fig.~\ref{5876} and below). Our principal aim in this paper is to understand this phenomenon via a deeper examination of the spectroscopy. 

In Fig.~\ref{profiles} we illustrate the profiles of selected spectral lines in the 2015 spectrum, shown in red. As the 2015 observation included only the '346' (3042-3872~\AA) and the '564' (4616-6653~\AA) settings of UVES, we employ the 2016 UVES observation (JD 2457644, obtained in the '437' setting, yielding wavelength coverage from 3752-4988~\AA) for the He~{\sc ii}\,$\lambda 4686$ line, which was located at the edge of an order in the 2015 spectrum. These data are shown in green. For comparison, we also show the April 2017 UVES spectrum (JD 2457867) for all lines (except H$\alpha$, because this line was saturated) and the two ESPaDOnS spectra \citep{2012MNRAS.419.2459W} selected to characterize the full range of short-term (rotational) variability of the spectrum observed in 2009/10. These are shown in black.


A significant change of both line position and shape is observed in essentially all lines in 2015-16, including both emission and absorption lines. In particular, we note that emission lines sensitive to the magnetosphere (i.e. those observed by \citealt{2008AJ....135.1946N} and \citealt{2012MNRAS.419.2459W} to vary significantly according to the 7.03\,d rotational period: Balmer lines, He~{\sc i} $\lambda 5876$, He~{\sc ii} $\lambda 4686$), emission lines insensitive to the magnetosphere (exhibiting weak to no detectable 7.03\,d variability: the $\lambda 4640$ complex of C~{\sc iii} and N~{\sc iii} lines, He~{\sc i} $\lambda 4921$), and photospheric absorption lines (N~{\sc iv} $\lambda 5200/04$, He~{\sc ii} $\lambda 5411$, C~{\sc iv} $\lambda 5801/11$) show significantly different morphologies in the 2015-16 spectra relative to the historical record as illustrated by the ESPaDOnS spectra. Moreover, the 2017 spectra show profiles that are generally more similar to those observed historically, supporting our impression that 2015 represented the extreme of the event, after which the spectrum has been gradually returning to normal.

In the remainder of this report, we focus our attention on one notable feature of the variability of \HD: the apparent wholesale shifts of the C~{\sc iii} and N~{\sc iii} $\lambda 4630-50$ emission peaks, and the $\lambda 5801/11$ absorption cores, in the profiles illustrated in Fig.~\ref{profiles}. We note in particular the fact that the absorption lines appear to shift in the direction { opposite} to that of the emission lines.

\section{Radial velocities of the C~{\sc iii}/N~{\sc iii} $\lambda 4634-47$ emission lines and the C~{\sc iv} $\lambda 5801/11$ absorption lines}

We have determined the radial velocities (RVs) of the N~{\sc iii}~$\lambda\lambda 4634.140, 4640.640$ and C~{\sc iii}~$\lambda 4647.418$ lines by measuring the wavelengths of the sharp peaks of the emission of these three lines in each spectrum. Due to blending, we did not measure the C~{\sc iii}~$\lambda 4650$ line RV. We computed the simple mean RV of the emission lines. These values are reported in the fourth column of Table~\ref{observations}. Similarly, we determined the radial velocities of the C~{\sc iv}~$\lambda\lambda 5801.310, 5811.970$ lines by measuring the wavelengths of the absorption cores of these two lines. Again, we computed the simple mean RV of these absorption lines. These values are reported in the fifth column of Table~\ref{observations}. Note that some spectra didn't cover one or the other of these regions, explaining blank entries in Table~\ref{observations}. 


\citet{1977ApJ...215..561C} reported radial velocities of a number of lines in the spectrum of \HD\ with typical uncertainties of order 5~\kms. While line-to-line velocity differences were reported, no significant velocity variations were reported by those authors, nor did they report RVs for the $\lambda 4640$ or $\lambda 5800$ lines. Similar results were reported by \citet{2008AJ....135.1946N}.

We re-measured the RVs of the UCLES spectrum and FEROS spectra discussed by \citet{2008AJ....135.1946N}, finding velocities in reasonable agreement with the values reported in those papers.


Finally, we also considered the SMARTS RVs discussed by \citet{2008AJ....135.1946N}. However, since these measurements are of lower precision, they didn't add significantly to our analysis. Therefore we will not discuss them any further here.

We checked for systematic errors in the RVs by verifying the absence of any detectable shifts of the strong, sharp Na~{\sc i} interstellar features located close to the C~{\sc iv} absorption lines. Random uncertainties were estimated for each line from the dispersion of measurements from independent spectra obtained close in time (when available), or from multiple measurements of individual spectra (when they were not). These values, typically $\sim 1-2$~km\,s$^{-1}\,$ for the $\lambda 4640$ emission lines and $\sim 2-4$~km\,s$^{-1}\,$ for the $\lambda 5800$ absorption lines, are reported in Table~\ref{observations}.



Both the $\lambda 4640$ and $\lambda 5800$ RVs exhibit significant changes between 1995 and 2017. These changes are much larger than the $\sim 2$~km\,s$^{-1}\,$ scatter of the RVs of these lines measured from the full collection of ESPaDOnS spectra of \citet{2012MNRAS.419.2459W}. We note in particular that (i) in accordance with our initial impression, the $\lambda 4640$ and $\lambda 5800$ RVs vary in opposition to one another, and (ii) the largest difference in RV between the $\lambda 4640$ and $\lambda 5800$ lines (approximately 54~km\,s$^{-1}\,$) occurred in 2015, at the time of the largest distortion of the He~{\sc i} $\lambda 5876$ line.

\begin{figure}
\begin{tabular}{ccc}
\includegraphics[width=5.4cm,angle=-90]{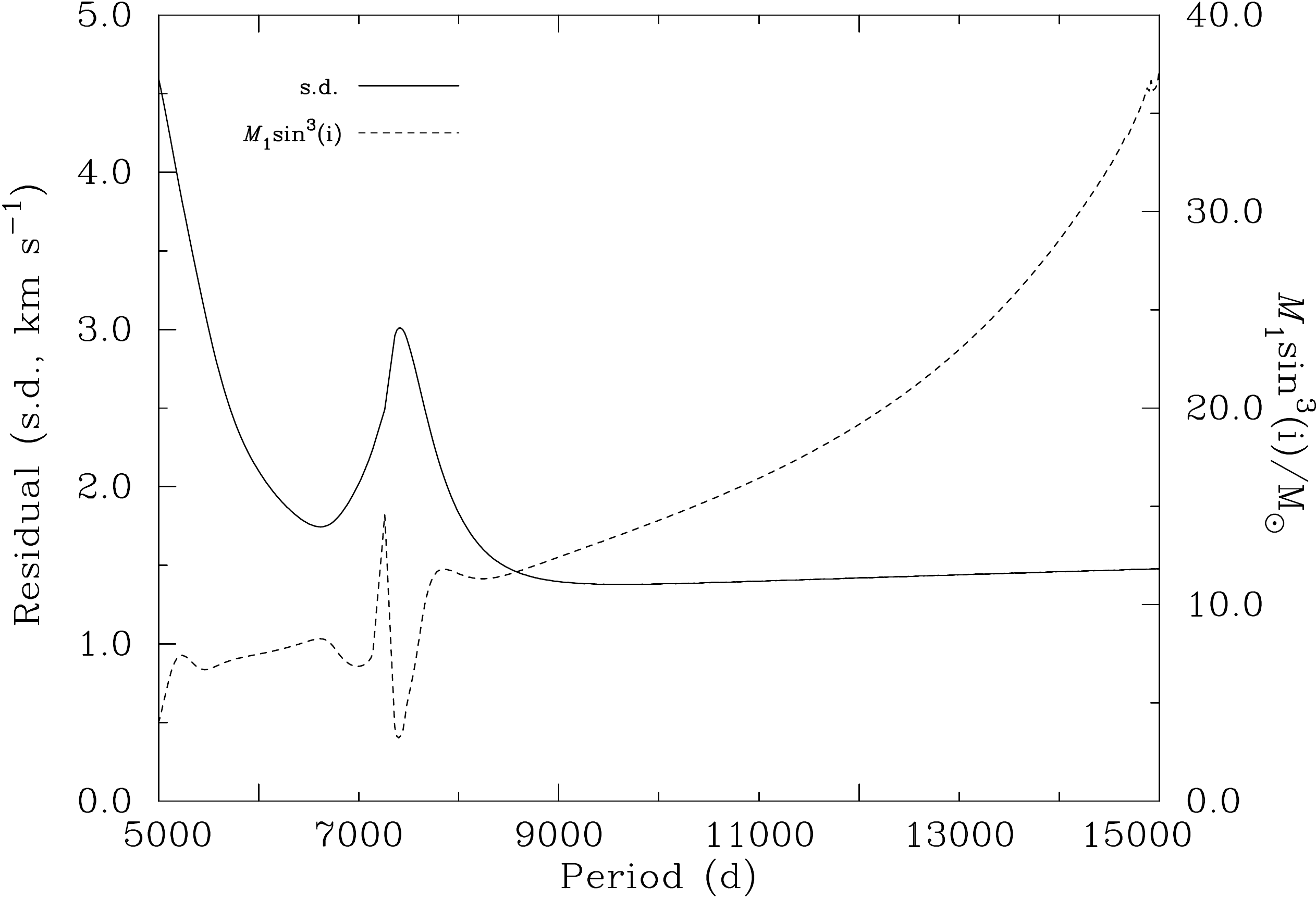} \\
\end{tabular}
\caption{{\em Solid curve -}\ Residuals from best-fit orbital solutions for a range of fixed period. {\em Dashed curve -}\ Implied projected mass of the primary star versus period. \label{period}}
\end{figure}


\begin{table}
\begin{centering}
\begin{tabular}{lrr}
\hline
Quantity & Long period & Short period \\
& solution & solution \\
\hline
P (d) &  $9591 \pm 350$ & $6617 \pm 50$\\
$T$ (d) & $2456991 \pm 20$ & $2457116\pm 35$\\
$K_1$ (km\,s$^{-1}\,$) & $17.3 \pm 2.0$ & $13.5\pm 1$\\
$K_2$ (km\,s$^{-1}\,$) & $25.2 \pm 2.5$ & $20.1 \pm 0.5$\\
$\gamma_1$ (km\,s$^{-1}\,$) & $-22.4 \pm 0.5$ & $-21.6 \pm 0.5$\\ 
$\gamma_2$ (km\,s$^{-1}\,$) & $-24.1 \pm 0.2$ & $-24.9 \pm 0.2$\\ 
$e$ & $0.75 \pm 0.03$ & $0.58 \pm 0.02$\\
$\omega$ ($\degr$) & $342 \pm 2$ & $347 \pm 3$\\
$M_1/M_2$ & $1.5 \pm 0.3$ & $1.5 \pm 0.2$\\
$M_1\sin^3 i$ ($M_\odot$) & $13.1\pm 5$& $8.5\pm 1$\\
$M_2\sin^3 i$ ($M_\odot$)& $9.0 \pm 4$& $5.7 \pm 1$\\
$a\sin i$ (AU) & $24.8\pm 3.5$ & $16.6 \pm 0.8$\\
\hline\hline
\end{tabular}
\caption{Orbital solution computed from the $\lambda 4640$ and $\lambda 5800$ RVs.\label{orbit_params}}
\end{centering}
\end{table}

\begin{figure*}
\begin{tabular}{ccc}
\hspace{-0.5cm}\includegraphics[width=9cm]{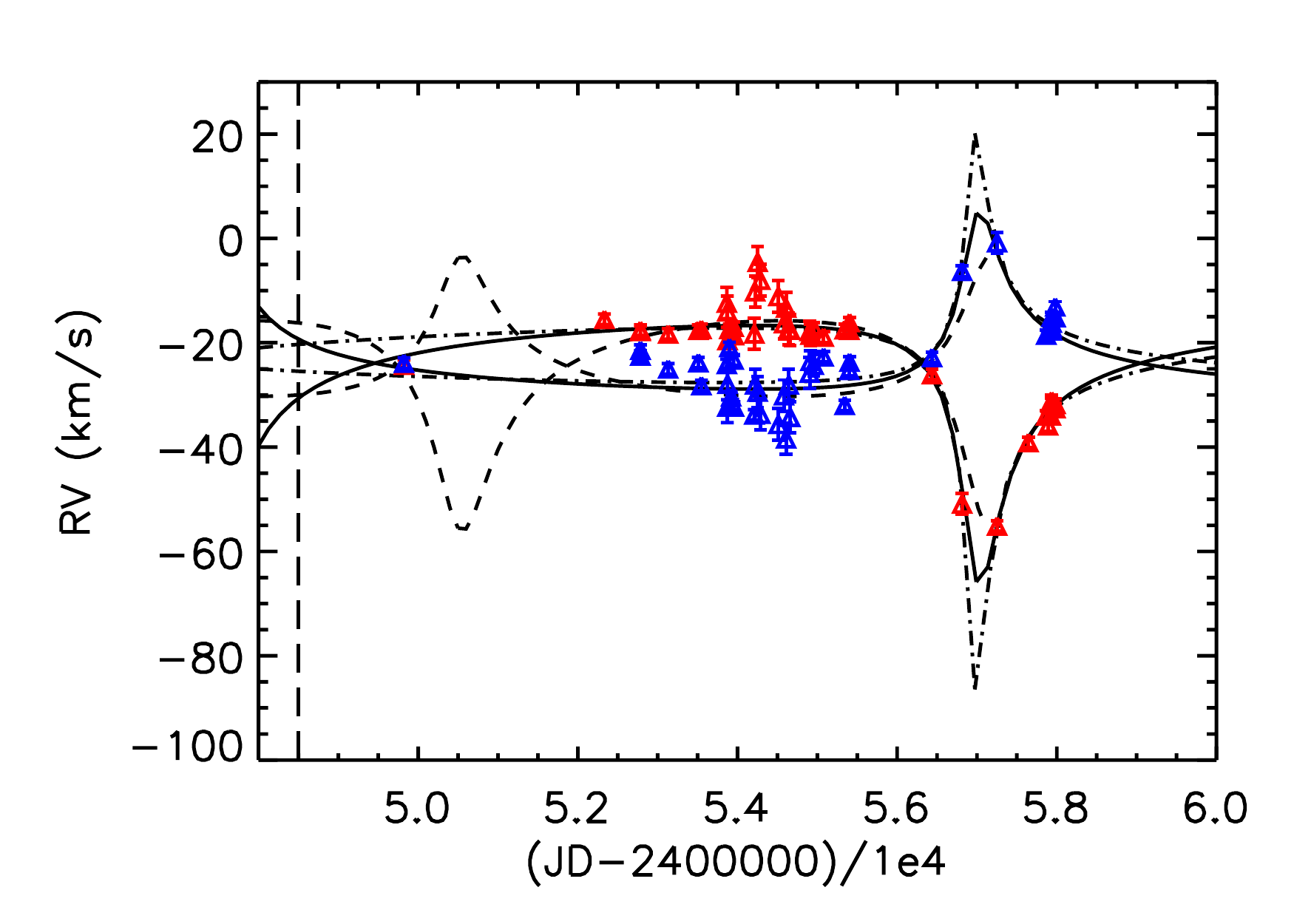}\includegraphics[width=9cm]{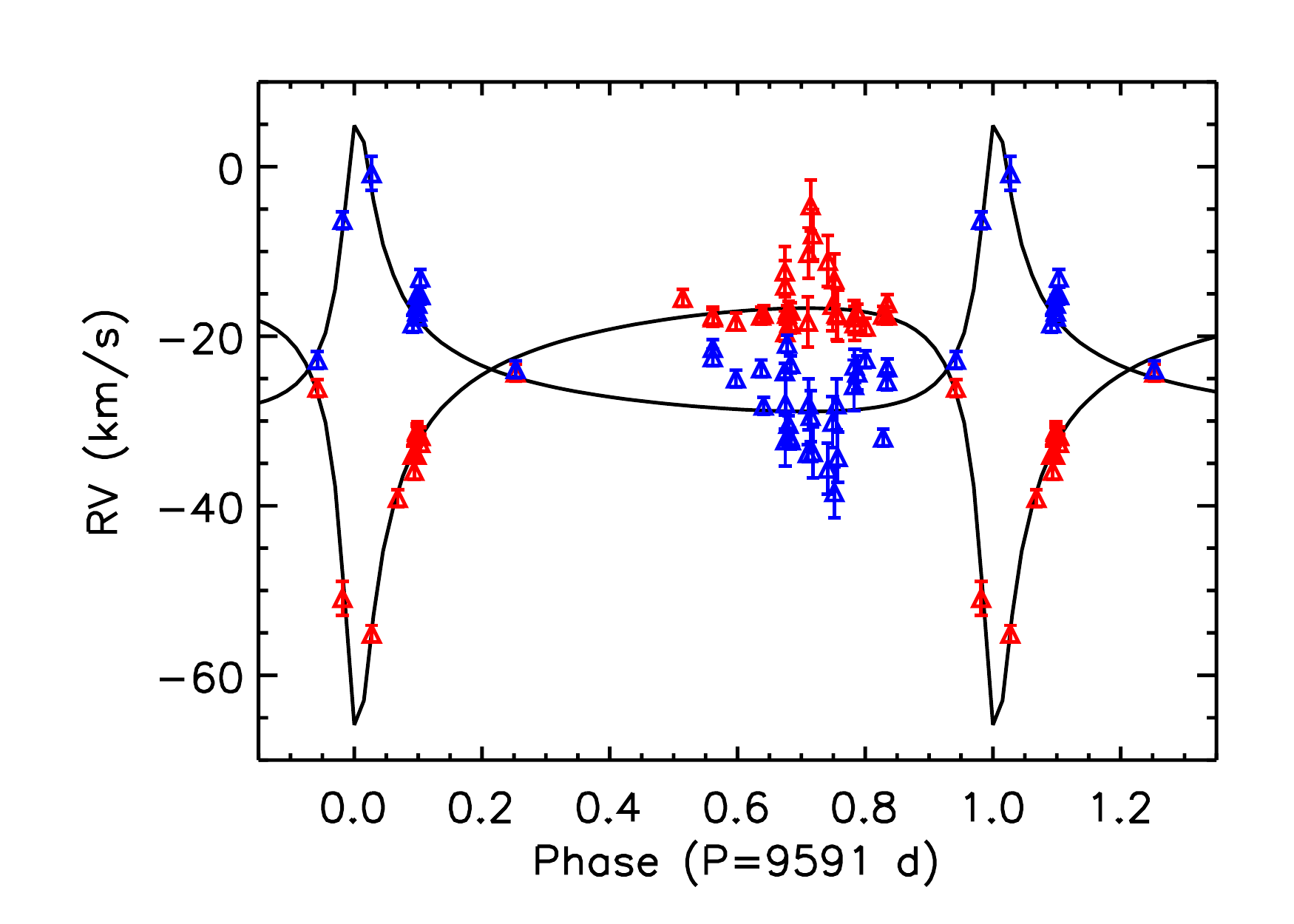} \\
\end{tabular}
\caption{\label{orbit}Model orbits fit to the RVs. {\em Left - }\ RV showing models versus JD corresponding to $P=9591$~d (the ``long-period solution", solid curve), $P=6617$~d (the ``short-period solution", dashed curve), and an arbitrary model with $P=13000$~d (a ``very-long period" solution, dot-dashed curve). {\em Right -}\ RVs phased and folded according to the $P=9591$~d period. In both plots the blue symbols are the $\lambda 5800$ measurements, corresponding to the primary component. The red symbols are the $\lambda 4640$ measurements, corresponding to the secondary star. The vertical dashed line in the left panel represents the date of the EMMI observation by Nota et al.}
\end{figure*}

\section{\HD\ appears to be a massive SB2}

The character of the RV measurements is highly suggestive of binary motion, in which the $\lambda 4640$ emission lines diagnose the RV of one stellar component, and the $\lambda 5800$ absorption lines diagnose the other component. 

To test this hypothesis, we employed the IDL orbital fitting code {\sc Xorbit} \citep{1992ASPC...32..573T} to model the orbit. This code determines the best-fitting orbital period $P_{\rm orb}$, Julian date of periastron passage ($T$), eccentricity ($e$), longitude of the periastron ($\omega$), semi-amplitudes of each component's radial velocities ($K_1$ and $K_2$), and the radial velocities of the centre of mass ($\gamma_{\rm 1, 2}$ for both components), performing least-squares fits to the measured radial velocities. 

An acceptable fit to the data is achieved with the orbital hypothesis. The spectroscopic data appear to cover only slightly more than one orbital period, and with only sparse sampling of the phase of periastron passage. As a consequence, the RVs allow two distinct periods to fit the data more or less equally well. As illustrated in Fig.~\ref{period}, the residuals from the best-fit orbital solutions for a range of fixed period (from 2000 to 20000~d at 20~d steps) confirm that there are reasonable solutions for periods around 6500~d (formally $6617\pm 50$~d) and $\gtrsim 9000$~d (with a formal best fit at $9591\pm 350$~d), with this longer period actually giving marginally better residuals. The ambiguity in $P$ arises essentially because we are as yet unable to determine if the first (UCLES) observation falls before or after the previous periastron passage. Nevertheless, the two periods yield similar orbital geometries, with roughly the same RV amplitude ratios and zero points. The eccentricity derived for the longer period is somewhat higher ($0.75$ versus $0.58$). While periods larger than $9000$~d are relatively unconstrained by the existing data, we note that the implied projected mass of the primary star becomes rapidly unphysical for periods longer than 10,000-11,000 days. This is also illustrated in Fig.~\ref{period}.

\citet{1996ApJS..102..383N} reported acquiring an ESO EMMI spectrum of \HD\ between 17-20 September, 1991. While the original data no longer appear to be available, illustrations of line profiles provided in their paper suggest that the $\lambda 5876$ line was likely experiencing distortions at that time similar to those exhibited in 2015/2017. In particular, the $\lambda 5876$ profile illustrated in their paper bears a strong similarity to those obtained in May 2013 and April/August 2017 (See Fig.~1). These observations bracket the epoch of strongest distortion of the line. Since the observation obtained by \citet{1996ApJS..102..383N} was obtained roughly 8500~d prior to the most recent periastron passage, if we suppose that the spectroscopic variability is associated with orbital phase (noting
that the appearance of line-profile anomalies in 2014/15 coincided
closely with periastron passage, regardless of the adopted orbital period) this would a imply period of $\sim 8000$~d (if the EMMI spectrum was obtained before the previous periastron passage) or $\sim 9500$ (if it was obtained after the previous periastron passage). The latter timescale is consistent with the long-period model, which we tentatively adopt as our reference model. 


Using {\sc Xorbit} we were able to achieve a reasonable simultaneous fit to the measured RVs of both sets of lines, for both orbital periods described above. Fig.~\ref{orbit} illustrates the RVs of both stars versus JD and phased assuming the longer period. The RV semi-amplitudes imply a mass ratio $M_1/M_2=1.5$, with an uncertainty of 15-20\%. The orbital parameters derived from the SB2 solution adopting the longer and shorter periods are summarised in Table~\ref{orbit_params}. Given the limited quality and coverage of the data, we focus on the general characteristics of the solution rather than the details.

\section{Discussion}

We have reported a remarkable change of the spectrum of the magnetic Of?p star \HD\ that departs qualitatively from the long historical record of the star's periodic 7.03\,d variability.
 
Our measurements of the $\lambda 4640$ (the N~{\sc iii}~$\lambda\lambda 4634, 4640$ and C~{\sc iii}~$\lambda 4647$ emission lines) and $\lambda 5800$ (the $\lambda\lambda 5801, 5811$ absorption lines) RVs show convincingly that the spectrum variation is significantly influenced by binary motion of two stellar components. If our orbital interpretation is correct, then it appears that the $\lambda 4640$ RVs are principally sensitive to the motion of one star (the lower-mass secondary), while the $\lambda 5800$ RVs trace the motion of the higher-mass primary star. Examination of the variations of the H$\alpha$ and He~{\sc ii}~$\lambda 4686$ profiles shown in Fig.~\ref{profiles} -- lines observed by \citet{2008AJ....135.1946N,2012MNRAS.419.2459W} to vary strongly according to the 7.03\,d period -- reveals that they also show bulk RV shifts qualitatively consistent with the $\lambda 4640$ lines. Since we expect that these emission lines arise principally in the magnetosphere, we tentatively interpret this to imply that the lower-mass secondary star is also the magnetic star. 

\subsection{Astrometric considerations}

As noted in Sect.~1, long baseline interferometric observations of \HD\ obtained with VLTI/PIONIER by \citet{{2014ApJS..215...15S}} revealed two equally bright components in the $H$-band ($\Delta H=0.00\pm 0.02$). Those authors assumed the Hipparcos distance to derive the instantaneous projected separation. However, the Hipparcos parallax they used is only significant at $3\sigma$ confidence. Given $V=6.7$ and $E(B-V)=0.61$ \citep{2017A&A...599A..61M}, adopting the Hipparcos distance the implied absolute magnitude of the system is $M_{\rm V}=-3.3$, which is somewhat faint for an O star (and another 0.75~mag fainter if we correct for a roughly equal-brightness companion).  

As discussed in the introduction, the recent Gaia DR2 parallax re-evaluation \citep{2018arXiv180409366L} implies a distance of $1.14\pm 0.06$~kpc. However, \citet{2018arXiv180409366L} explicitly discuss DR2 results for unresolved/partially resolved multiple systems. In particular, they note: ``In this release, all sources beyond the solar system are still treated as single stars, that is, as point objects whose motions can be described by the basic five-parameter model. For unresolved binaries (separation $\ltsim 100$\, mas), the photocentre is consistently observed and the astrometric parameters thus refer to the position and motion of the photocentre in the wavelength band of the $G$ magnitude. Orbital motion and photometric variability may bias the astrometric parameters for such sources." Given the results of \citet{{2014ApJS..215...15S}} reporting the detection of a companion of similar $H$-band magnitude at a separation of about 21~mas, we interpret the DR2 results with a degree of skepticism.

Given the weak constraints from the Hipparcos parallax, and our concerns about the reliability of the Gaia parallax, it may well be more secure to adopt a distance based on assumed membership of Ara OB1 \citep[e.g.][which lists $m-M=10.7$, implying $d=1.4$~kpc]{1978ApJS...38..309H}. We note that Mahy et al. quote a distance of 1.3 kpc, while \citet{2009MNRAS.400..518M} give 1.1 and 2.8 kpc for Ara ``OB1A" and ``OB1B", respectively.  Distances of order 1~kpc are more consistent with the fluxes expected of typical O stars, and also more consistent with the Gaia parallax (notwithstanding the caveat noted above). It therefore seems plausible that the distance to \HD\ is similar to that implied by the Gaia parallax, and considerably larger than implied by the Hipparcos parallax (i.e. 1.0-1.4~kpc, rather than 425~pc). This conclusion is also supported by the interstellar Ca~{\sc ii} column density reported by \citet{2006MNRAS.367.1478H} interpreted using the relationships of \citet{2015A&A...582A..59S}.


%

Gaia DR2 also reports a $G$=12.1 (as compared to $G$=6.6 for \HD) source located at an angular separation of 3.3 arcsec. This source certainly corresponds to the $\Delta H$=5.4 neighbour observed by \citet{2014ApJS..215...15S}. The parallax of this source yields a distance somewhat smaller than (but within $2\sigma$ of) the brighter source. However, given the caveat discussed above, it is not clear that we can draw any significant conclusions regarding its physical association with the main system. Certainly the formal parallaxes are not very different from one another.




Knowing the spectroscopic orbit, we can in principle calculate the projection of the orbit onto the plane of the sky, given two additional angles: the position angle of the ascending node, and the orbital inclination \citep[e.g.][]{1996A&A...310..235H}.  The former is of no particular interest (it merely defines the orientation of the projected orbit), but we can readily plot the projected linear separation of the components as a function of orbital phase for any given inclination.

\begin{figure*}
\begin{tabular}{ccc}
\includegraphics[width=5.9cm,angle=-90]{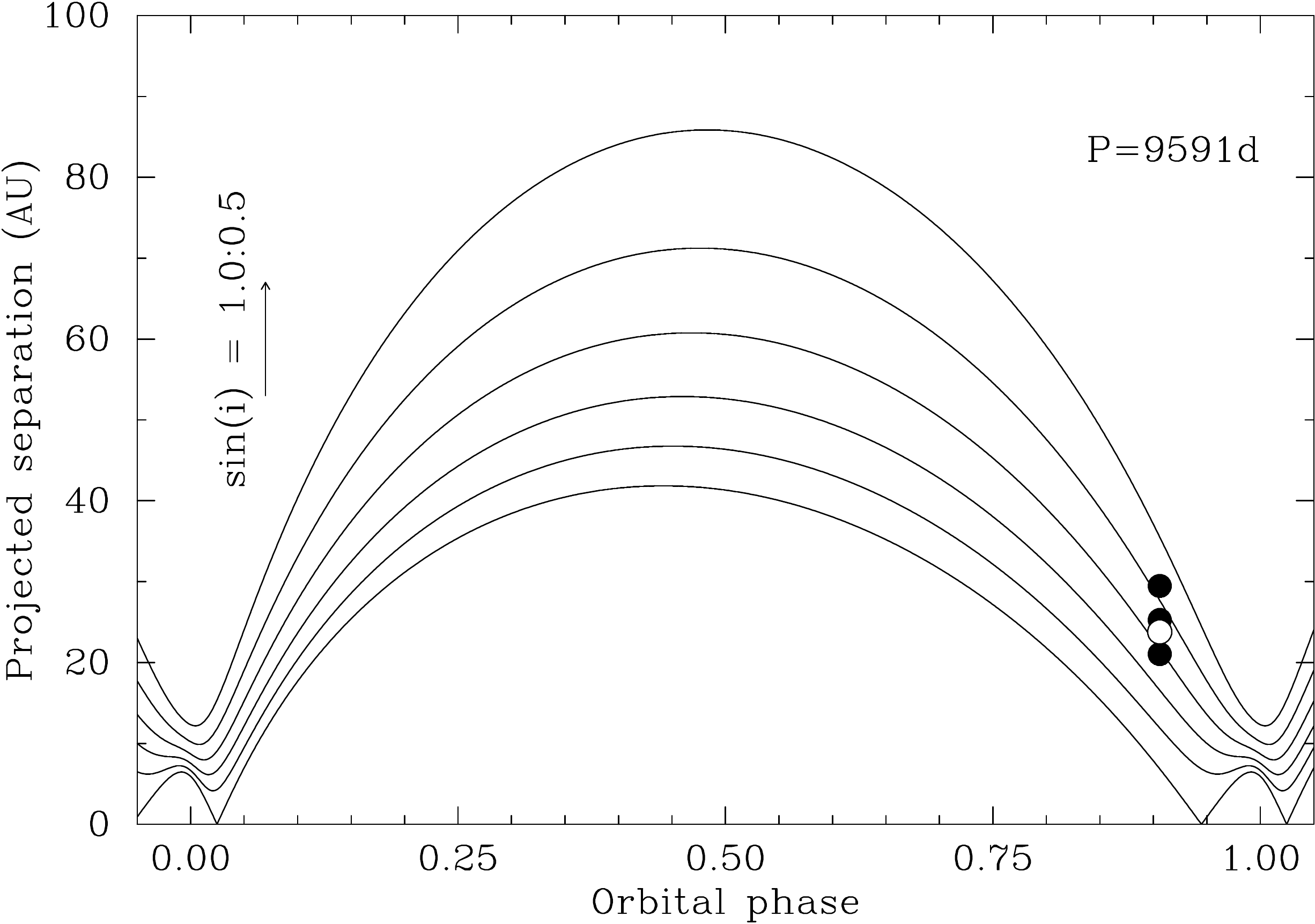}\hspace{0.5cm} \includegraphics[width=5.9cm,angle=-90]{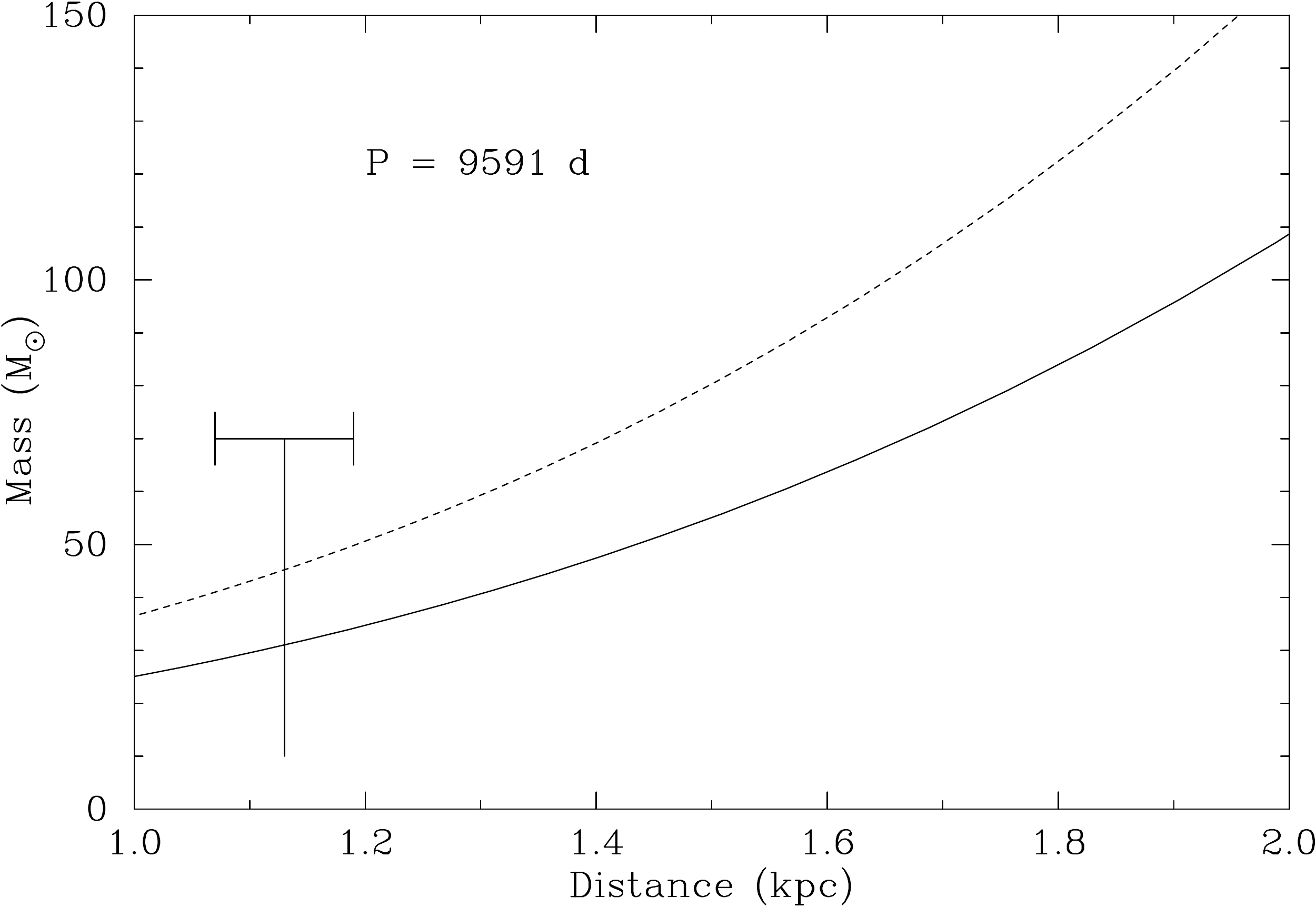} \\
\end{tabular}
\caption{{\em Left -}\ Projected separation of the components of \HD\ assuming the longer-period ($P=9591$~d) orbital solution. The curves correspond to inclinations $\sin i = 1.0, 0.9, 0.8$... (from the bottom up). Filled circles are the separation for assumed distances of 1.0, 1.2, and 1.4 kpc (bottom to top). The open circle represents the Gaia DR2 distance, for reference. {\em Right -}\ Constraint that the single SMASH+ observation provides on masses as a function of distance. The two curves correspond to $M_1$ and $M_2$. The Gaia DR2 distance is indicated by the vertical line. We remind the reader that the Gaia distance may be systematically affected by the presence of the close interferometric companion.\label{separation}}
\end{figure*}

Fig.~\ref{separation} (left panel) presents such plots for the longer orbital period;  the curves are for sin(i) = 1.0, 0.9, 0.8... (from the bottom up). At the moment, we only have one direct measurement of the angular separation from SMASH+;  we take $\rho=21.05$~mas at JD 2456088. Converting the date to phase assuming the longer period, and the angular measure to linear measure, we can add this observation to the plots -- shown as dots for assumed distances of 1.0, 1.2, and 1.4 kpc (bottom to top). Because the curves in these plots are not parallel, additional astrometric observations can, in principle, constrain distance and inclination separately.  But even this single observation has some utility, as it constrains $\sin i$ to $0.65\pm 0.1$ for this plausible range of distance. 

Adopting $\sin i=0.65\pm 0.1$ (or $i=40\pm 8\degr$), the implied masses of the components are $49~M_\odot$ and $34~M_\odot$ for the primary and secondary. As illustrated in Fig.~\ref{separation} (right panel), the full range of permitted masses - given the uncertainty on both distance and period - for the secondary is rather large. (We note that Wade et al. (2012) reported a mass of \HD\ of $60~M_\odot$, but without benefit of knowledge of the system's multiplicity.) The other known Of?p stars have masses ranging from $30-45~M_\odot$ \citep[e.g.][]{2015ASPC..494...30W}, in agreement with the result for the secondary, but consistent with the mass of either star given the uncertainties. 


\subsection{Spectrum formation and composition}

Why do the $\lambda 4640$ and $\lambda 5800$ features seem to distinguish the motion of the stars so clearly? It is known that the N~{\sc iii}~$\lambda\lambda 4634/40/42$ and C~{\sc iii}~$\lambda\lambda 4647/50/52$ emissions are strongly peaked at spectral type O5 in Ofc stars \citep{2010ApJ...711L.143W}. This spectral type seems too early given the secondary's mass, but strong and variable C~{\sc iii}~$\lambda\lambda 4647/50/52$ emission is also a defining characteristic of the magnetic Of?p stars. {Moreover, examining spectra in our possession reveals that the N~{\sc iii} and C~{\sc iii} emission line intensities of \HD\ are the strongest of any of the Galactic Of?p stars, and stronger than those of HD\,191612 (which has the next-strongest emission in this region) by about 80\%. } It therefore seems reasonable to assume that the magnetic secondary's emission lines dominate in the $\lambda 4640$ region.


EWs of the C~{\sc iv} $\lambda\lambda 5801/11$ lines as a function of spectral type \citep{1996ApJS..103..475F} peak at O6-O7.5 for all luminosity classes. The combination of a mid-O star (strong C~{\sc iv}) plus a late-O star (weak C~{\sc iv}) of similar continuum brightness is roughly consistent with the observed EW of the $\lambda 5801$ line. Such a scenario could potentially explain the sensitivity of the C~{\sc iv} lines to the primary's motion. {We note that this necessarily implies that the more massive primary star is hotter than the magnetic secondary star. As a consequence, the identical $H$-band magnitudes of the two stars need to be understood. Non-rotating evolutionary tracks of \citet{ekstrom2012} for 35 and 50~$M_\odot$ stars differ by somewhat less than a factor of $\sim 2$ (i.e. 0.3 dex) in bolometric luminosity through most of the main sequence, but knowledge of the effective temperatures of the two stars and their relative radii would be necessary to evaluate the expected $H$-band flux ratio. This puzzling characteristic of the system should be investigated further once additional constraints on the masses, radii, and temperatures of the stars have been obtained.}

Given that the details of the formation of the spectrum are poorly understood, it is worth evaluating if all of the observed variability could potentially be explained by the intrinsic variability of a single star. While the evidence presented above leading us to interpret the change in the spectrum of \HD\ as a consequence of the mutual orbital motion of two massive O-type stars is compelling, it might be argued that other phenomena, e.g. asymmetric infilling of spectral lines results in the {appearance} of RV variations. However, we underscore that photospheric absorption lines (high ions of C and N with weak sensitivity to the magnetosphere and circumstellar environment), emission lines that are clearly sensitive to the magnetosphere (showing strong modulation according to the 7.03~d rotational period of the magnetic star), as well as emission lines that are not modulated by the rotational period, are all impacted by this episode. Qualitatively, this seems hard to explain by, e.g. variable emission resulting from enhanced mass loss. Moreover, careful examination of the C~{\sc iv} $\lambda 5801$ line (in particular) shown in Fig. 2 shows that its variability does not seem to be consistent with infilling by emission; the observed RV variation represents a wholesale shift of the entire line. Similarly, the $\lambda 4640$ lines experience no significant change in shape, but are rather moved (fully and in tandem) to the blue in Fig. 2. Finally, the independent interferometric observation provides clear corroborating evidence of binarity in the detection of a companion of similar magnitude.  We therefore conclude that a model invoking intrinsic line profile variations of a single star to explain these diverse phenomena is far less simple, and therefore less preferable, than the adopted eccentric binary interpretation, which provides a straightforward interpretation of the data, yields physical parameters that are broadly consistent with the expected properties of the components, and is quantitatively falsifiable.  As a consequence, while the detailed formation of the system's spectrum clearly deserves further investigation, we conclude that the binary origin of the apparent RV variations is currently the best interpretation of the observed variability.

\subsection{X-rays}

The new X-ray observations were acquired to investigate if the X-ray emission of \HD\ has been impacted by the phenomena responsible for the recent spectroscopic changes. 

To avoid cross-correlation problems, we compare the fitting results described in Sect.~2.2 to those derived from fits to the previous Chandra datasets \citep{2012ApJ...746..142N}. The spectral parameters are very similar, except for a slight increase in normalization factors (e.g. for the ``cool" solution, the first normalization remains constant but the second one is 20\% larger). The observed flux is consequently larger, by $\sim$20\%, and this flux difference appears significant (6$\sigma$). Compared to the flux derived from older XMM data \citep{2008AJ....135.1946N}, the change is even smaller ($\sim 5$\%). 

No large change in the X-ray flux of \HD\ is expected in the context of magnetically-confined wind shocks, since the small inferred inclination angle of the star leads to the magnetosphere being viewed from a similar perspective at all phases, yielding weak rotational modulation. Hence the 20\% increase in flux compared to the previous Chandra observations likely requires a different explanation. In massive binaries, X-rays are sometimes emitted due to a strong collision between the stellar winds. In the case of long periods, as is the case here, such collisions are adiabatic in nature, and the X-ray luminosity varies as $1/{\rm separation}$ with absorption effects only playing a role close to the periastron passage (for a review, see \citealt{2016AdSpR..58..761R}). The three X-ray datasets were taken at JD= 2451966.2 (XMM, \citealt{2008AJ....135.1946N}), 2455353.6 (Chandra; \citealt{2012ApJ...746..142N}), and 2458286.8 (new Chandra data), which correspond to relative separations of 1.7, 1.2, and 1.1, respectively, for the long-period solution. If colliding winds alone were responsible for the X-ray emission of \HD, then the largest flux would have been expected for the new observation and the lowest (35\% lower flux) for the XMM observation, with little change between the two Chandra observations, but the observations indicate otherwise.


For the short-period solution, relative separations would be 1.2, 1.3, and 1.1 for the three datasets, respectively, with expected increases in flux of $\sim$8\% and $\sim$15\% for the new dataset compared to the XMM and old Chandra observations, respectively. At first sight, the observed X-ray properties may thus appear very compatible with the short-period scenario. However, the good agreement in amplitude changes must be considered in the context that the colliding wind contribution is small far from periastron (see e.g. \citealt{2012A&A...546A..37N}). For the first Chandra observation, the X-ray flux should be close to the canonical relation $\log(L_{\rm X}/L_{\rm BOL})\sim -7$ naturally expected for massive stars, whereas $\log(L_{\rm X}/L_{\rm BOL})\sim -6.25$ is derived from data. This suggests that the bulk of the X-ray emission does not come from colliding winds. Indeed, the emission level observed previously is fully compatible with predictions of confined winds models \citep{2014ApJS..215...10N}. This implies that any colliding wind contribution will be diluted, leading to lower variability levels. Considering this fact, the X-ray variability of \HD\ is thus not fully explained, but is likely not incompatible with a contribution from colliding winds in a binary system.


\subsection{Other outstanding issues, and future observations}

\begin{figure}
\begin{tabular}{ccc}
\includegraphics[width=8cm]{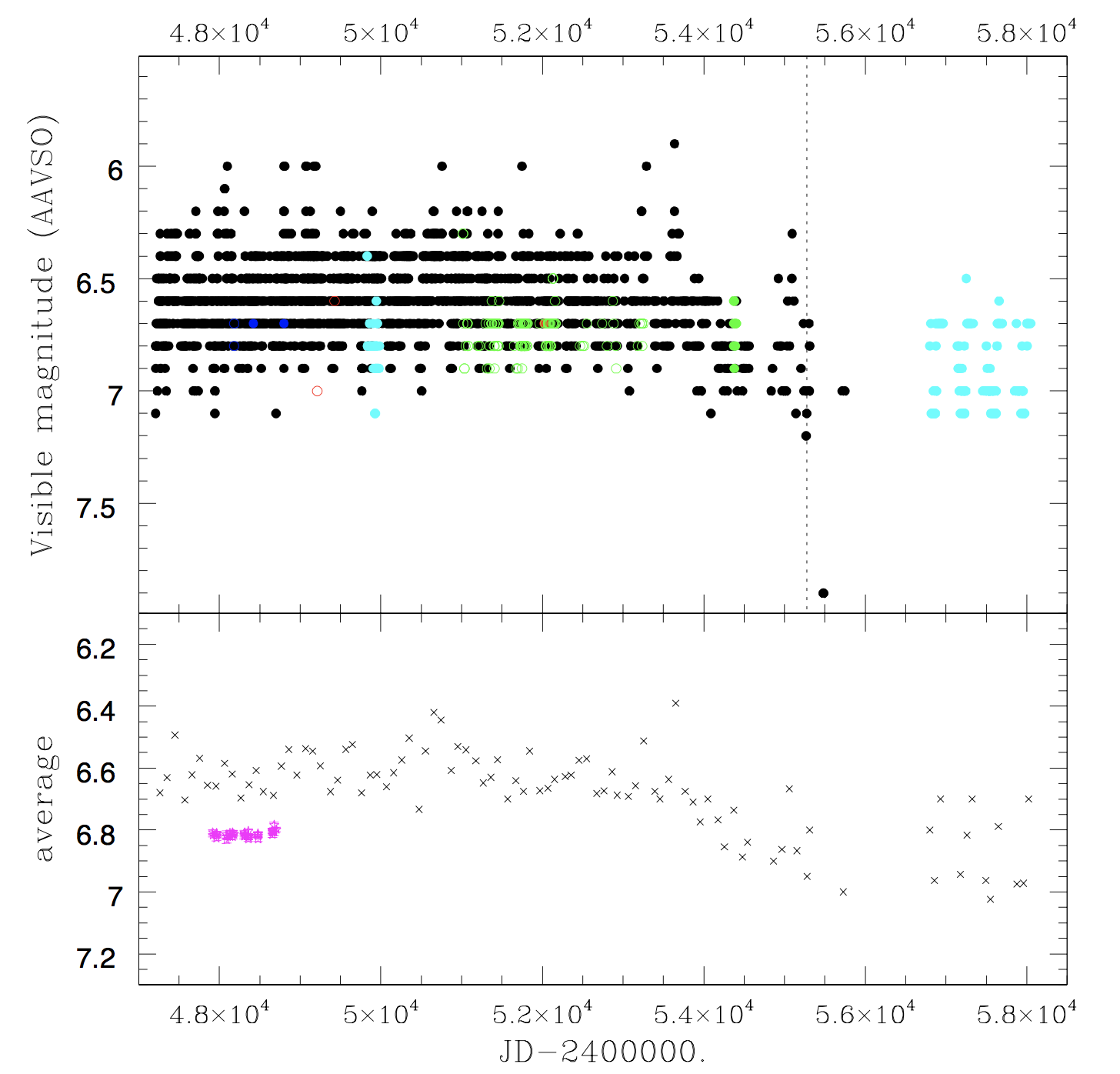} \\
\end{tabular}
\caption{Visual magnitude measurements made by the RASNZ, with different observers coded by different symbols. The vertical dotted line indicates when comparison stars were changed. The lower frame shows the same data binned in 100-day averages (in grey), along with the Hipparcos photometry (in purple).\label{aavso}}
\end{figure}

Spectral variability of Of?p stars is often accompanied by photometric changes \citep[e.g.][]{2004ApJ...617L..61W}. However, photometry is scarce for \HD. \citet{2008AJ....135.1946N} reviewed published analyses of photometric observations of \HD, summarizing evidence for long-term variability and performing an analysis of the Hipparcos and Tycho photometry. They concluded that no significant change or periodicity is present (see as well figure 13 of \citealt{2012MNRAS.419.2459W}). In addition to the photometry discussed by those authors, ASAS-3 photometry of \HD\ was obtained between 2000-2008. However, the star is in the saturation regime of the survey, and the measurements show clear evidence of saturation-related systematics. Another existing source of photometric measurements is the AAVSO\footnote{https://www.aavso.org/} database, with 2550 points since February 1988 reported by the Royal Astronomical Society of New Zealand (RASNZ, Fig. \ref{aavso}). Unfortunately, except for five points, all values are only pre-validated; in addition, there is a gap in measurements between mid-2011 and mid-2014, an epoch of great interest in the context of this paper. Nevertheless, even if scatter is quite large, there seems to be a decline in brightness, by $\sim$0.2~mag, since the end of 2006 (JD=2454000). The data were obtained by 8 different observers: a single one (``JA") is responsible for 85\% of them, but another one (``WPX") obtained all recent ones. However, the decline is also seen in the sole ``JA" dataset. Given the limited quality of the existing photometry, it is clear that little can be concluded apart from the fact that the star has exhibited no large photometric variations (i.e. greater than a few tenths of magnitudes) between 1988 and 2014. The AAVSO photometry of \HD\ (along with the Hipparcos photometry) as a function of Julian date are illustrated in Fig.~\ref{aavso}.


In the Introduction, we speculated that the unexpected variability may have its origin in the probable binarity reported by \citet{2014ApJS..215...15S}. Based on our spectroscopic monitoring, we tentatively conclude that an important component of the change in the spectrum of \HD\ is associated with variable distortion by a companion comparable in mass and luminosity to the magnetic star, in an eccentric, long-period (26~year) orbit. For the longer-period solution, the inferred semi-major axis and eccentricity imply a closest approach (around JD 2456991, the beginning of December 2014) of the two stars of about 9.7~AU,  equal to roughly 140 times the radius of a typical Of?p star. This would suggest that tidal interactions between the two components are unlikely, even at periastron. Nevertheless, both spectroscopy and unpublished interferometry (Sana, priv. comm.) agree that the greatest distortion of the spectrum occurs when the stars are physically closest to one another. We also point out that the eccentricity could potentially be larger, given that the values obtained from our fits are derived from relatively few RV points somewhat removed from the exact date of periastron passage. 

As pointed out by \citet{2017A&A...599A..61M}, HD 148937 is the only magnetic O-type star known to be surrounded by a nebula that can be attributed to mass loss. $\theta^1$~Ori C and NGC\, 1624-2 have surrounding nebulosities, but they are H~{\sc ii} regions associated with the star-forming environments of these stars. The exact formation process of this nebula raises many questions. Mahy et al. suggest that multiple epochs of nebular ejection have occurred, $1.2-1.3$~Myr ago, and a more recent event $0.6$~Myr ago. They estimate the mass of the ionized gas in the nebula to be $\sim12.4~M_\odot$, but this estimate is uncertain. They discuss evolutionary scenarios in detail, but only from the perspective of a single star: a giant eruption (an LBV-like event?) triggered by the stellar wind and the magnetic field, or a merger event between two massive stars in a binary configuration. The merger scenario focuses on explaining the magnetic field of the secondary star, but arguments against it work for both secondary and primary. Such a merger would require that one of the two existing O-type stars would have previously been two lower-mass stars in a tight orbit, and then through subsequent stellar and/or dynamical evolution the stars would have merged. Given that the pre-merger masses of these stars would have been substantially smaller than that of the current secondary or primary, they should have been less evolved. So the mechanism by which such a merger would have occurred is unclear. 

\citet{2012MNRAS.419.2459W} report that, compared to other lines of similar strength and Land\'e factor, the He~{\sc i}\, $\lambda 5876$ line exhibits an anomalously strong Stokes $V$ signature. In addition, the star appears to show a Stokes $V$ profile given the star's relatively weak magnetic field. The $V$ profile appears to be somewhat wider than the sharper principal component of its Stokes $I$ LSD profile \citep[see Fig. 7 of][]{2012MNRAS.419.2459W}. It seems plausible that both of these characteristics could be a result of the SB2 nature of the system, with the Zeeman signature contributed by one component, but the Stokes $I$ profile defined by contributions of both stars. Alternatively, emission might infill the blue wing of the profile. Since these spectra were obtained before the recent episode of spectral evolution began, such emission would most likely be of magnetospheric origin.


{It will be very interesting to eventually determine the orientation of the binary orbit and its relationship to the nebular geometry. This will help to clarify if the binary system governs the nebular geometry, and if/to what event the magnetic field may also contribute. We note that the derived orbital inclination is in reasonable agreement with the spectroscopic constraint that the rotational axis inclination of the magnetic star be less than about $30\degr$. As a consequence, aligned spin and orbital axes cannot be ruled out. And it may well be that both binarity and magnetism contribute to the observed nebular characteristics, e.g. binarity as the origin, with the magnetic field sculpting the ``figure 8" shape described by \citet{2017A&A...599A..61M}.} Moreover, wide-field imagery of the ISM and nebulosity adjacent to \HD\ \citep[see apod.nasa.gov/apod/ap160330.html and the analysis by][]{2006romero} may be indicative of larger-scale and longer-timescale interactions of the system.

Could the (likely important) contribution of the presumably non-magnetic primary star to the spectrum be related to the surprising cycle-to-cycle scatter in the 7.03\,d EW curves reported by \citet{2012MNRAS.419.2459W}? If the primary star exhibits episodic mass ejections (as might be inferred from the presence of the nebula), it may be stochastically variable on shorter timescales as well. More generally, the presence of a bright second star similar in spectral type to the magnetic star has implications for the quantitative interpretation of the stellar spectrum, and of the magnetic measurements. Another, more exotic possibility is that the variability of the magnetospheric lines really does probe intrinsic variations of the magnetospheric structure. In that case, ascribing the recent variability to a significant reconfiguration of the Of?p star's magnetosphere might be plausible.However, all of this is speculation, and more detailed investigation of the spectral contributions of the two stars - before and during the 2015 event - will help to answer these questions and understand the detailed origins of the spectral evolution. 

The discovery of the likely binary nature of \HD\ is significant, as it contributes to the ongoing discussion about the relationship of binarity to massive-star magnetism \citep[e.g.][]{2015IAUS..307..330A}. Nevertheless, our binary solution is not unique, and it is subject to the limitations of incomplete phase coverage and limited data quality. Dedicated spectroscopic monitoring should be continued, with a view to eventually observing the next periastron in either December, 2032 (shorter period) or February, 2039 (longer period). Continued interferometry should ultimately provide an independent test of the orbital model proposed here, and ultimately combining interferometry and RVs should securely determine the orbital geometry and stellar masses. In addition, future spectroscopic or spectropolarimetric monitoring covering the 7.03\,d period will test if the magnetospheric configuration (and potentially more remarkably the rotational period) of the Of?p star has changed.

Simultaneously, a detailed analysis of the spectra of the components must be performed, including a fine determination of the stellar properties. This should allow evaluation if simple shifting of two spectra in velocity is sufficient to explain the observed variation, or if intrinsic variability must be invoked. The magnetic and magnetospheric properties of the Of?p star should be re-evaluated taking into account the spectral contamination of the companion. Additionally, monitoring on the rotational timescale could serve as an immediate check if there has been any change in magnetospheric variability.

 \section*{Acknowledgments}
We dedicate this work to our friend and colleague Nolan R. Walborn, who passed away during the final stages of this investigation.

We thank the anonymous referee for their careful review and thoughtful comments.
 
This work has made use of data from the European Space Agency (ESA) mission
{\it Gaia} (\url{https://www.cosmos.esa.int/gaia}), processed by the {\it Gaia}
Data Processing and Analysis Consortium (DPAC,
\url{https://www.cosmos.esa.int/web/gaia/dpac/consortium}). Funding for the DPAC
has been provided by national institutions, in particular the institutions
participating in the {\it Gaia} Multilateral Agreement.
 
GAW acknowledges Discovery Grant support from the Natural Sciences and Engineering Research Council of Canada.  YN acknowledges support from the Fonds National de la Recherche Scientifique (Belgium), the Communaut\'e Fran\c caise de Belgique, and the PRODEX XMM contract. RG is Visiting Astronomer, Complejo Astron\'omico de la Rep\'ublica Argentina and the National Universities of La Plata, C\'ordoba and San Juan. RB and JA thank the DIDULS projects PR18143 and PR16142, respectively. JC and AF acknowledge support from an NSERC Discovery Grant and a SERB
Accelerator Award from Western University.  UCLES data from April 1995 was provided by Simon O Toole at AAT. We exploited data from ESO programs 075.D-0369(A), 077.D-0705(A) , 082.C-0446(B), 082.D-0136(A), 194.C-0833(D), 194.C-0833, 266.D-5655(A). The authors acknowledge the contribution of H. Sana, who provided helpful discussion and comments, access to unpublished interferometric results, and kind support to GAW during a research visit to KU Leuven. 
\bibliography{article}

\end{document}